\documentclass[twocolumn]{aastex63}

\usepackage[normalem]{ulem}
\usepackage{footnote}
\usepackage{subfigure}
\usepackage{xspace}
\usepackage[utf8]{inputenc}
\usepackage{multirow,amsmath}
\usepackage{float}
\usepackage{apjfonts}

\newcommand\msun{{M}$_{\odot}$\xspace}                   

\newcommand{\code}[1]{\texttt{#1}}
\newcommand{\mesa}{\code{MESA} }
\newcommand{\MESA}{\mesa}
\newcommand{\flash}{\code{FLASH} }
\newcommand{\FLASH}{\flash}

\shorttitle{3D Massive Star Models}
\shortauthors{Fields \& Couch}

\submitjournal{ApJ}

\begin{document}
\title{\Large Three-Dimensional Hydrodynamic Simulations of Convective Nuclear Burning In Massive Stars Near Iron Core Collapse}

\correspondingauthor{C.~E.~Fields}
\email{carlnotsagan@lanl.gov}

\author[0000-0002-8925-057X]{C.~E.~Fields}
\altaffiliation{RPF Distinguished Fellow}
\affiliation{Center for Theoretical Astrophysics, Los Alamos National Laboratory, Los Alamos, NM 87545, USA}
\affiliation{Computer, Computational, and Statistical Sciences Division, Los Alamos National Laboratory, Los Alamos, NM 87545, USA}
\affiliation{X Computational Physics Division, Los Alamos National Laboratory, Los Alamos, NM 87545, USA}

\author[0000-0002-5080-5996]{Sean M.~Couch}
\affiliation{Department of Physics and Astronomy, Michigan State University, East Lansing, MI 48824, USA}
\affiliation{Department of Computational Mathematics, Science, and Engineering, Michigan State University, East Lansing, MI 48824, USA}
\affiliation{Facility for Rare Isotope Beams, Michigan State University, East Lansing, MI 48824, USA}

\begin{abstract}
Non-spherical structure in massive stars at the point of iron core collapse
can have a qualitative impact on the properties of the ensuing core-collapse supernova explosions 
and the multi-messenger signals they produce. 
Strong perturbations can 
aid successful explosions by strengthening turbulence in the post-shock region. 
Here, we report on a set of $4\pi$ 3D hydrodynamic simulations of O- and Si-shell burning in massive star models of varied initial masses 
using \texttt{MESA} and the \texttt{FLASH} simulation framework.
We evolve four separate 3D models for roughly the final ten minutes prior to, 
and including, iron core collapse. 
We consider initial 1D \texttt{MESA} models with masses of 14-, 20-, and 25 \msun 
to survey a range of O/Si shell density and compositional configurations. We characterize the
convective shells in our 3D models and compare them to the corresponding 1D models. 
In general, we find that the angle-average convective speeds in our 3D simulations near collapse are \emph{three} to \emph{four} times larger than the convective speeds predicted by \texttt{MESA} at the same epoch for our chosen mixing length parameter of $\alpha_{\rm{MLT}}=1.5$. 
In three of our simulations, we observe significant power in the spherical harmonic 
decomposition of the radial velocity field at harmonic indices of $\ell=1-3$ near collapse.
Our results suggest that large-scale modes are common in massive stars near collapse and 
should be considered a key aspect of pre-supernova progenitor models.
\end{abstract} 


\section{Introduction} 
\label{sec:intro}
Three-dimensional (3D) simulations of core-collapse supernova (CCSN) explosions have benefitted 
from imposing pre-supernova velocity perturbations that aim to replicate nuclear burning in Si- and 
O-shell convective regions. The inclusion of these perturbations were shown to lead to 
larger non-radial kinetic energy in the gain region providing turbulent pressure behind the 
stalled shock capable of driving explosion in a model that otherwise failed to explode without 
perturbations \citep{couch_2013_aa}. In the work of \citet{oconnor_2018_ab}, they 
impose perturbations in the 3D CCSN explosion of a 20 \msun model \citep{farmer_2016_aa}.
Their 3D perturbed CCSN models evolved closer towards shock runaway and explosion but 
they did not observe shock runaway in any of the eight 3D simulations performed. The 
pre-supernova perturbations in the Si-shell region lead to an increase in the gravitational 
wave (GW) amplitude at $t_{\textrm{pb}}\approx 200$ ms over a frequency band of 
200 - 1000 Hz. This result suggests that convective perturbations can also lead to qualitative 
differences in the multi-messenger signals produced in CCSN simulations. 

Work by \citet{couch_2015_aa} presented the results of a 3D CCSN progenitor model 
evolved for the final $\approx$ 155 s prior to and including iron core-collapse. Using this 
3D progenitor model they performed two CCSN explosion simulations using the 3D 
progenitor model and a 1D angle-average of the 3D model. They
found that increased turbulent motions in the post shock region in the 3D progenitor explosion model 
can aid in successful explosion.
This model also showed a slight increase in turbulent kinetic energy in the
gain region, a similar result to the CCSN models with artificial perturbations in \citet{couch_2013_aa}. 
More recently, \citet{muller:2017} used the 18 \msun 4$\pi$ 3D progenitor of \citet{muller_2016_aa} to 
perform three 3D CCSN explosion models. In their 3D explosion model using a 
3D progenitor they found an increasing diagnostic explosion energy and baryonic mass 
of the PNS values closer agreement with those expected from observations \citep{pejcha_2015_ab}. 
The two
additional CCSN explosion models in their study - one a reduced velocity field 3D progenitor and 
the other a 1D angle-average initial model were less energetic and the 1D progenitor model
failed to explode. 
The results of this study suggest that 3D progenitors can aide in closing the gap between low explosion
energy and PNS properties predicted by other works using 1D progenitors \citep{burrows_2020_aa}.

Multidimensional CCSN progenitor models have been performed recently 
\citep{muller_2016_aa,Yoshida_2019,fields_2020_aa,yadav_2019_aa,yoshida_2021_aa}. Many of 
these simulations show convective properties that suggest favorable impact on the 
neutrino-driven CCSN explosion mechanism. In \citet{yadav_2019_aa} they observe 
large-scale mixing due to merger of the O- and Ne-shells, a results which lead to
large radial mach numbers in the merged shell regions. Despite this progress, it is expected 
that the convective properties in massive star will span a range of strength and flow
dynamics over the initial mass range for CCSNe \citep{muller_2015_ab}. Currently, 
only a handful of 3D simulations sample this mass range and provide predictions 
for the Si- and O-shell convective properties of 3D massive star models. 

In this Paper, we build on previous efforts exploring 3D progenitor models in the moments 
prior to collapse. We perform a total of four $4\pi$ 3D hydrodynamic simulations of Si- and 
O-shell burning for up to the final ten minutes prior to and including gravitational instability 
and iron core-collapse. We evolve models of initial  $M_{\textrm{ZAMS}}$ =14-, 20, and 25 \msun 
using the \texttt{FLASH} simulation framework and the 1D stellar evolution code, \texttt{MESA} 
\citep{fryxell_2000_aa,paxton_2011_aa,paxton_2013_aa,paxton_2015_aa,paxton_2018_aa}.
This work is novel because: (1) - we present four 3D long-term hydrodynamic simulations of 
O/Si shell burning in multiple progenitors, (2) - we investigate the impact of initial perturbations 
in pre-supernova hydrodynamic simulations in two 3D simulations of a 20 \msun model, and (3)
we compare the convective properties of our 3D models to the predictions of Mixing Length Theory (\texttt{MLT}) 
in three different initial 1D progenitor models. 

This paper is organized as follows. In \S~\ref{sec:methods} we describe our computational methods and
initial 1D \texttt{MESA} progenitor models. In \S~\ref{sec:results_all} we present the results 
of our 3D simulations including characterizing their global properties, 
comparing stellar properties to the 1D \MESA models,
and exploring turbulent entrainment in the O-shell in our 20 \msun models.
Lastly, in \S~\ref{sec:discussion} we summarize our main findings.  

\section{Computational Methods and Initial Models} 
\label{sec:methods}
Our methods follow those of \citet{fields_2020_aa} (referred to as FC20). We draw an initial 
1D progenitor for mapping into 3D to simulate the final minutes of Si- and O-shell burning towards iron core-collapse. 
Here, we highlight the difference in our initial conditions and the initial progenitor set chosen for this study. 
The \texttt{MESA} inlists, initial and final 1D \texttt{MESA} models, and the four 3D progenitor models 
produced as a part of this work are publicly available \href{https://zenodo.org/record/4895094#.YOYPwhNKjzg}{online}.

\subsection{1D \MESA Stellar Evolution Models}
We employ the stellar evolution toolkit, Modules for Stellar Astrophysics (\texttt{MESA}-revision 12115)
\citep{paxton_2011_aa,paxton_2013_aa,paxton_2015_aa,paxton_2018_aa,paxton_2019_aa}, for 
our spherically-symmetric 1D models. In total, we evolve three solar metallicity, zero-age 
main-sequence (ZAMS) mass progenitors: 14\msun, 20\msun, and 25\msun. Each progenitor is 
evolved in \texttt{MESA} from the pre-MS to a time approximately \emph{10 minutes} prior to iron core-collapse. 
These models utilize temporal/spatial parameters from previous studies 
shown to provide adequate converge in core quantities at the level of uncertainty due to network size and
reaction rates \citep{farmer_2016_aa,fields_2018_aa}. Our 1D models use the same approximate network as used in FC20, an $\alpha$-chain network 
that follows 21 isotopes from $^{1}$H to $^{56}$Cr \citep{timmes_2000_ab}. Our \MESA models 
are non-rotating and do not include magnetic fields. Mass loss is included using the `\texttt{Dutch}` wind scheme 
with an efficiency value of $\eta_{\texttt{Dutch}}=$0.8. Mixing processes and efficiency values are the 
same as used in FC20, which use a mixing length parameter of $\alpha_{\rm{MLT}}=1.5$ in all convective regions.

\begin{deluxetable*}{ccccccccccc}
\renewcommand{\arraystretch}{1.0}
\tablecolumns{10}
\tablewidth{2.0\linewidth}
\tablecaption{Properties of the initial progenitor models at the time of mapping}
\tablehead{
\hline
\colhead{Initial Mass} & 
\colhead{$\xi_{2.5}$} & 
\colhead{$r_{\rm{low,Si}}$} &  \colhead{$r_{\rm{high,Si}}$} & \colhead{$C_{\rm{Si}}$} & \colhead{$\mathcal{M}_{\rm{rad.,Si}}$} &
\colhead{$r_{\rm{low,O}}$} &  \colhead{$r_{\rm{high,O}}$} & \colhead{$C_{\rm{O}}$} & \colhead{$\mathcal{M}_{\rm{rad.,O}}$} &
\colhead{Simulation Time} \\
\colhead{($M_{\odot}$)} &  &
\colhead{(km)} & \colhead{(km)}  & \colhead{(g s$^{-1}$)}  &  &
\colhead{(km)} & \colhead{(km)}  & \colhead{(g s$^{-1}$)}  &  & 
\colhead{(sec)} 
}
\startdata
$\texttt{14m} $ & 0.016 &
2400 & 3000 & 1.2 $\times$ 10$^{28}$ & 3.0 $\times$ 10$^{-3}$ &
3130 & 15820 & 8 $\times$ 10$^{27}$ & 1.6 $\times$ 10$^{-2}$ &
530.71  \\
$\texttt{20m} $ & 0.151 &
2366 & 2854  & 5 $\times$ 10$^{27}$ & 1.0 $\times$ 10$^{-3}$ &
3120 & 42000 & 7 $\times$ 10$^{27}$ & 5.0 $\times$ 10$^{-3}$ &
643.83  \\
$\texttt{25m} $ & 0.519 &
3700 & 4150 & 6 $\times$ 10$^{27}$ & 8.0 $\times$ 10$^{-4}$ &
5500 & 44000 & 5.25 $\times$ 10$^{27}$ & 3.2 $\times$ 10$^{-3}$ &
606.95  \\
\enddata
\tablecomments{
3D \texttt{FLASH} simulation properties of the initial progenitor models at the time of mapping including the initial mass, 
compactness at a mass coordinate of $m=2.5$ \msun, O/Si-shell radial limits, the scaling factor used to produce a 
1 or 5$\%$ convective velocity profile according to \texttt{MESA}, the resulting radial mach number due to the perturbations, 
and the total simulation time.}
\label{tbl:initial_models}
\end{deluxetable*}

\subsection{3D \FLASH Hydrodynamic Stellar Models}
We simulate a total of four $4\pi$ 3D hydrodynamic models using the \texttt{FLASH} simulation framework \citep{fryxell_2000_aa,dubey_2009_aa}. Our models solve the equations of compressible hydrodynamics 
using the directionally unsplit piecewise parabolic
method (PPM), third-order spatial accuracy, solver implemented in \texttt{FLASH} \citep{lee_2008_aa}. We employ 
an HLLC Riemann solver \citep{toro_1999_aa} and use a Courant factor of 0.8. Self gravity is included assuming 
a monopole ($\ell=0$) gravitational potential \citep{couch_2013_ab}. Our domain extends to 100,000 km from the 
origin along an axis in Cartesian geometry utilizing the same boundary conditions as in FC20. 
Each model uses adaptive mesh refinement (AMR) with up to eight 
levels of refinement. We discuss our grid resolution in more detail in \S~\ref{sec:app1}.

The 3D models are initialized with perturbations in the Si- and O-shell region that are informed by their 1D 
\texttt{MESA} counterpart at the time of mapping. We use the same notation as in FC20, also used in \citet{muller_2015_ab} 
and \citet{oconnor_2018_aa}. In Table~\ref{tbl:initial_models} we show properties of our initial progenitor models at the time 
of mapping into \FLASH including the compactness ($\xi_{2.5}$), shell radii, simulation time, perturbation scaling factor $C$, 
and the resulting radial mach number ($\mathcal{M}_{\rm{rad.}}=v_{\rm{rad.}}/c_{\rm{s}}$, where $c_{\rm{c}}$ is the 
local sound speed) at the start 
of the simulation caused by the perturbations. The imposed 
perturbations are performed in the $r$ and $\theta$ components of the velocity field with topology determined 
by spherical harmonic indices and a scaling factor informed by the convective velocity profile of the 1D 
\texttt{MESA} model. 

For the 14 $M_{\odot}$ model we take the Si-shell region to be 
at a location of $\approx$ 2400 km to 3000 km within this region, a scaling factor $C$ of 1.2 $\times$ 10$^{28}$ g s$^{-1}$. 
The O-shell region is taken to be from $\approx$ 3130 km to 15820 km and we choose a
value of $C$ of  8.0 $\times$ 10$^{27}$ g s$^{-1}$. The locations of the shell regions were determined by the 
composition profiles and corresponded to radii enclosing the region at which the isotope was the most abundant and had a 
non-zero convective velocity according to \texttt{MESA}. For the Si-shell, the $C$ value was chosen to represent
the average value needed to produce 1$\%$ of the convective velocity predicted by \texttt{MESA}. 
In other words, we compute an approximate scaling factor such that the initial angle-average $\texttt{FLASH}$ 
convective velocity profile is equal 1\% of the mean convective speed in the 1D \texttt{MESA} model in this region. 
In the O-shell 
region the value of $C$ was chosen in a similar way except corresponding to 5$\%$. The Si-shell region 
used spherical harmonic and radial numbers of $n=1,\ell=9,m=5$ while the O-shell region used 
$n=1,\ell=7,m=5$ (initially larger scale perturbations). The resulting average radial mach number in the Si- 
and O-shell regions due to these perturbations were $\mathcal{M}_{\rm{rad.,Si}}\approx 3 \times 10^{-3}$ and $\mathcal{M}_{\rm{rad.,O}}\approx 1.6 \times 10^{-2}$, respectively. We note that for this model, we observe 
a weakly burning Ne-shell region at $\approx11000$ km. We do not employ separate initial perturbations for this shell.
Instead, our O-shell perturbations cover both of these shell regions in radius.

Our 20 \msun model was initialized in a similar fashion except the scaling factors chosen for both the Si- and 
O-shell regions corresponded to a value of $1\%$ of the average value needed to reproduce the convective 
velocity speeds predicted by \texttt{MESA}. In this model, the Si-shell region is 
located $\approx$ 2366 km to 2854 km within this region, a scaling factor $C$ of 5.0 $\times$ 10$^{27}$ g s$^{-1}$. 
The O-shell region is located at $\approx$ 3120 km to 42000 km and we choose a
value of $C$ of  7.0 $\times$ 10$^{27}$ g s$^{-1}$. For this particular model, a non-convective predominantly 
silicon region exists between these two regions. This model uses the same perturbation shape parameters.
In this model, the initial perturbations produce average radial mach numbers in the Si- 
and O-shell regions of $\mathcal{M}_{\rm{rad.,Si}}\approx 1 \times 10^{-3}$ and 
$\mathcal{M}_{\rm{rad.,O}}\approx 5 \times 10^{-3}$, respectively. For this particular model, this 
results in radial velocity speeds of $\approx \pm$ 40 km s$^{-1}$ in the O-shell region with speeds
in the Si-shell $\approx \pm$ 20 km s$^{-1}$

Lastly, the 25\msun model has a Si-shell region from $\approx$ 3700 km to 4150 km where we apply a scaling
factor of $C$ of 6 $\times$ 10$^{27}$ g s$^{-1}$. The O-shell region for this model extends from 5500 km to 44000 km
where we use an average scale factor of $C$ of 5.25 $\times$ 10$^{27}$ g s$^{-1}$. Similar to the 20 \msun 
model these scalings were chosen to reproduce approximately $1\%$ of the convective velocity predicted by
\MESA at the time of mapping. This model uses the same spherical shape parameters as the 20\msun model as well. 
In this model, the perturbations produce mach numbers in the SI- and O-shell regions of 
$\mathcal{M}_{\rm{rad.,Si}}\approx 8.0 \times 10^{-4}$ and 
$\mathcal{M}_{\rm{rad.,O}}\approx 3.2 \times 10^{-3}$, respectively.

Our \FLASH simulations utilize the approximate 21 isotope network ($\texttt{approx21}$) with
the same updated weak reaction rate used for electron capture onto $^{56}$Ni from \citet{langanke_2000_aa}. 
The Helmholtz stellar equation of state (EoS) as implemented in \texttt{FLASH} is used in all of 
our 3D simulations \citep{timmes_2000_aa}. 
We do not artificially enhance the total electron capture rates in any simulations presented here. 
All of our 3D simulations utilize a similar methods as in FC20 where we produce a 2D table from 
the \MESA profile data for the inner 1000 km from the point of mapping into \FLASH until iron core-collapse. 
Lagrange linear interpolation is then performed in time and radius to obtain a solution \FLASH models 
without the need for a call to the nuclear reaction network. This mapping provides a time-dependent 
inner boundary condition that ensures the model follows the central evolution of the \MESA model 
but is still significantly below the regions of interest for our study of the multi-D hydrodynamic properties. 
To help reduce artificial transient during mapping, we use the methods of \citet{zingale_2002_aa} in which we 
remap the 1D \MESA models to a new uniform grid with four times higher resolution than the finest grid spacing. 
We then alter the density profile to enforce hydrostatic equilibrium (HSE) and close the system with a call to the 
equation of state.

\begin{figure}[!t]
\centering{\includegraphics[width=1.0\columnwidth]{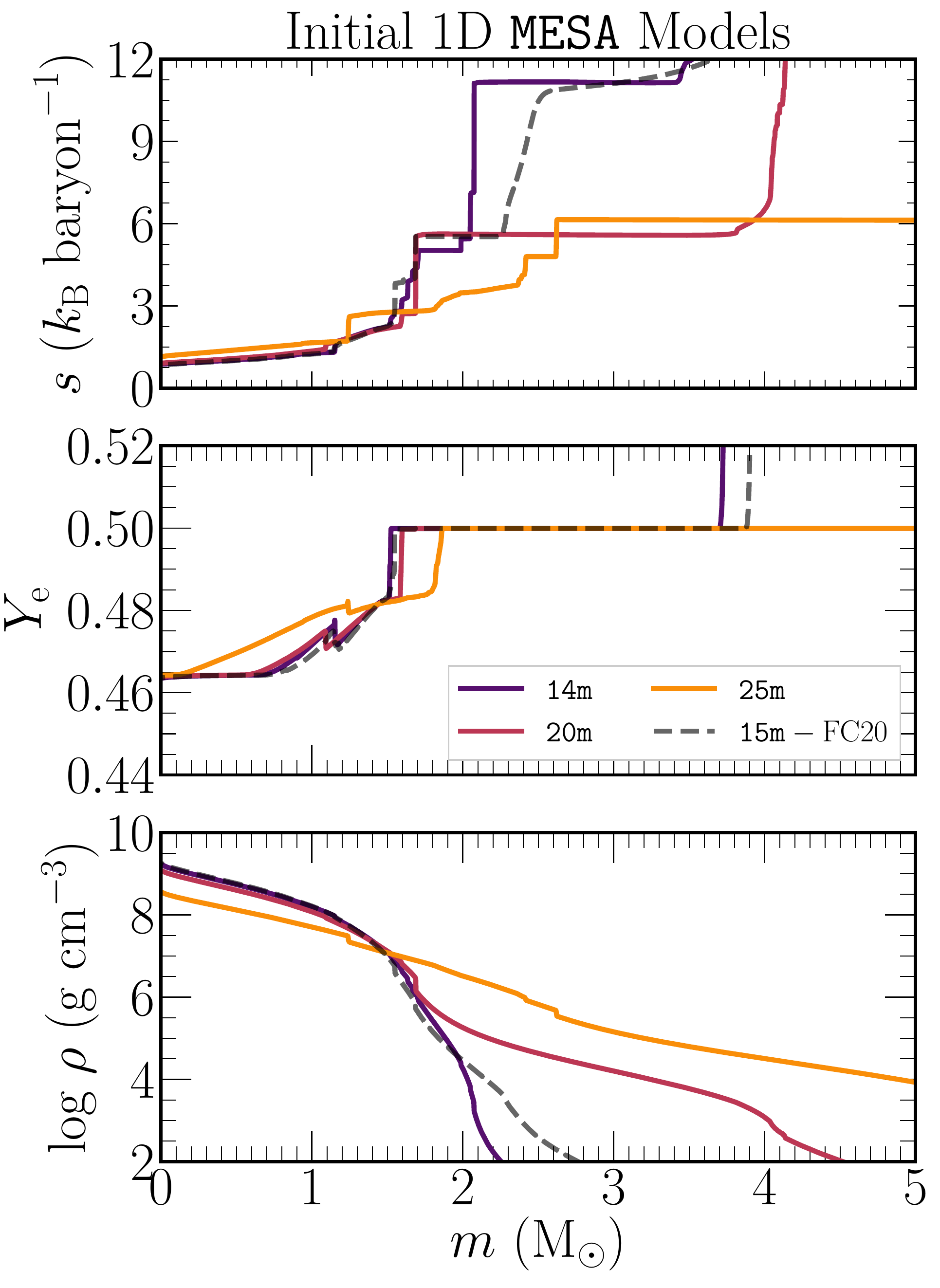}}
\caption{
Profiles of the specific entropy (top), electron fraction (middle), and density (bottom) for the 
three 1D \texttt{MESA} models at time of mapping into \texttt{FLASH}. Also shown is the 15 \msun
progenitor model from FC20 denoted by the gray dashed line.
}\label{fig:1d_structure}
\end{figure}

\begin{figure}[!t]
\centering{\includegraphics[width=1.0\columnwidth]{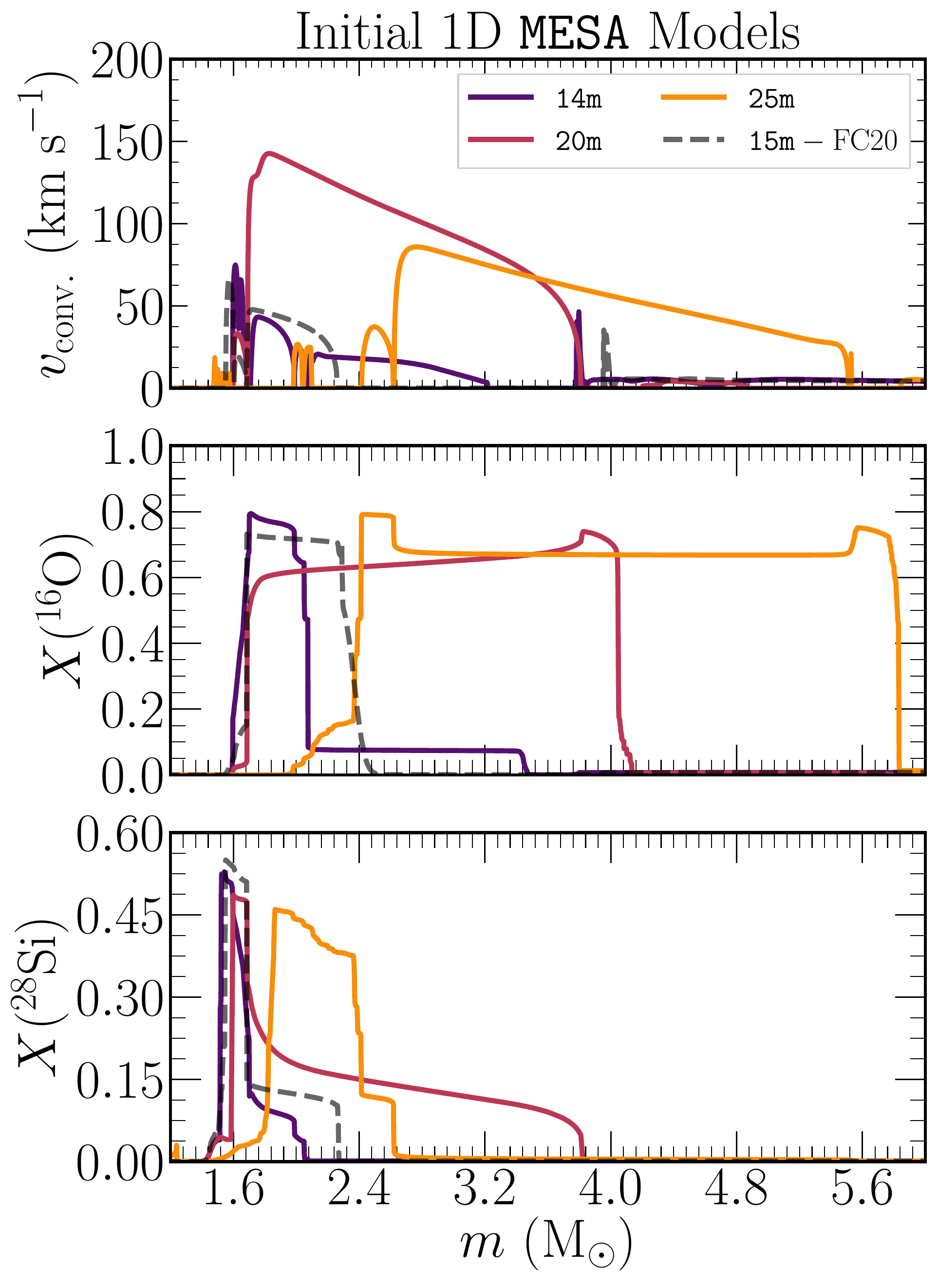}}
\caption{
Profiles of the convective velocity (top), oxygen-16 mass fraction (middle), and silicon-28 mass fraction (bottom) 
for the 1D \texttt{MESA} models at the same time as Figure~\ref{fig:1d_structure}.
}\label{fig:1d_isos}
\end{figure}

\subsection{Progenitor Models}
In this study, we aim to explore the range in hydrodynamic properties observed in pre-SN models 
in their evolution towards iron core-collapse. A metric commonly used to predict the outcome of a pre-SN models is the 
compactness parameter, 
\begin{equation}
\xi_{m} = \left.\frac{m / M_{\odot}}{R(M_{\rm{bary}}=m) / 1000 \textup{ km}} \right\vert_{t=t_{\rm{cc}}}~.
\label{eqn:xi}
\end{equation}
This quantity has been used to determine the outcome: explosion vs. implosion, in a range of different progenitors  \citep{oconnor_2011_aa,sukhbold_2016_aa}. In general, it has been shown that a lower compactness value favors
explosion while a higher value $\xi_{2.5}>0.45$ can result in failed explosion and formation of stellar mass 
black hole. The compactness parameter has also been shown to correlate with the integrated total 
neutrino emission count in successfully exploding model \citep{warren_2020_aa}. Models with larger 
$\xi_{2.5}$ were found to produce more to neutrinos owing to the more massive baryon proto-neutron 
star (PNS) mass associated to its larger value.

To sample the range of compactness seen in other studies we choose our initial 1D \MESA to span a 
range of values of compactness at time of mapping: $\xi_{2.5}\approx$ 0.016, 0.151, and 0.519, for the 14 \msun, 20 \msun, 
and 25 \msun models, respectively. For comparison, the 15 \msun model of FC20 had a core compactness 
of $\xi_{2.5}\approx0.014$. \citet{oconnor_2011_aa} predict that models with $\xi_{2.5}\geq0.45$ such as our
25 \msun would fail to explode (assuming a moderately stiff EoS such as LS220 \citep{lattimer_1991_aa}) forming a 
BH within $\approx$ 0.5 s post bounce. It should be noted that the core compactness values quoted in 
Table~\ref{tbl:initial_models}
should be viewed as a lower limit as we measure this quantity at the time of mapping into \FLASH and
not at core-collapse.

In Figure~\ref{fig:1d_structure} we show the specific entropy (top), 
electron fraction (middle), and density (bottom) for the 1D \texttt{MESA} models at 
time of mapping into \texttt{FLASH}. For comparison, we also plot the 15  \msun progenitor
from FC20. The 25 \msun 
represents the most shallow density profile likely contributing to its larger compactness value.
The 14 \msun density is most similar to that of the 15 \msun model of FC20. The 20 \msun
has a density profile whose shallowness is somewhat in between the 14 \msun and 25 \msun 
model. This is consistent with the trend seen for the values of compactness for these models.

\begin{figure}[!t]
         \centering  
        \begin{subfigure}{
                \includegraphics[width=0.47\textwidth]{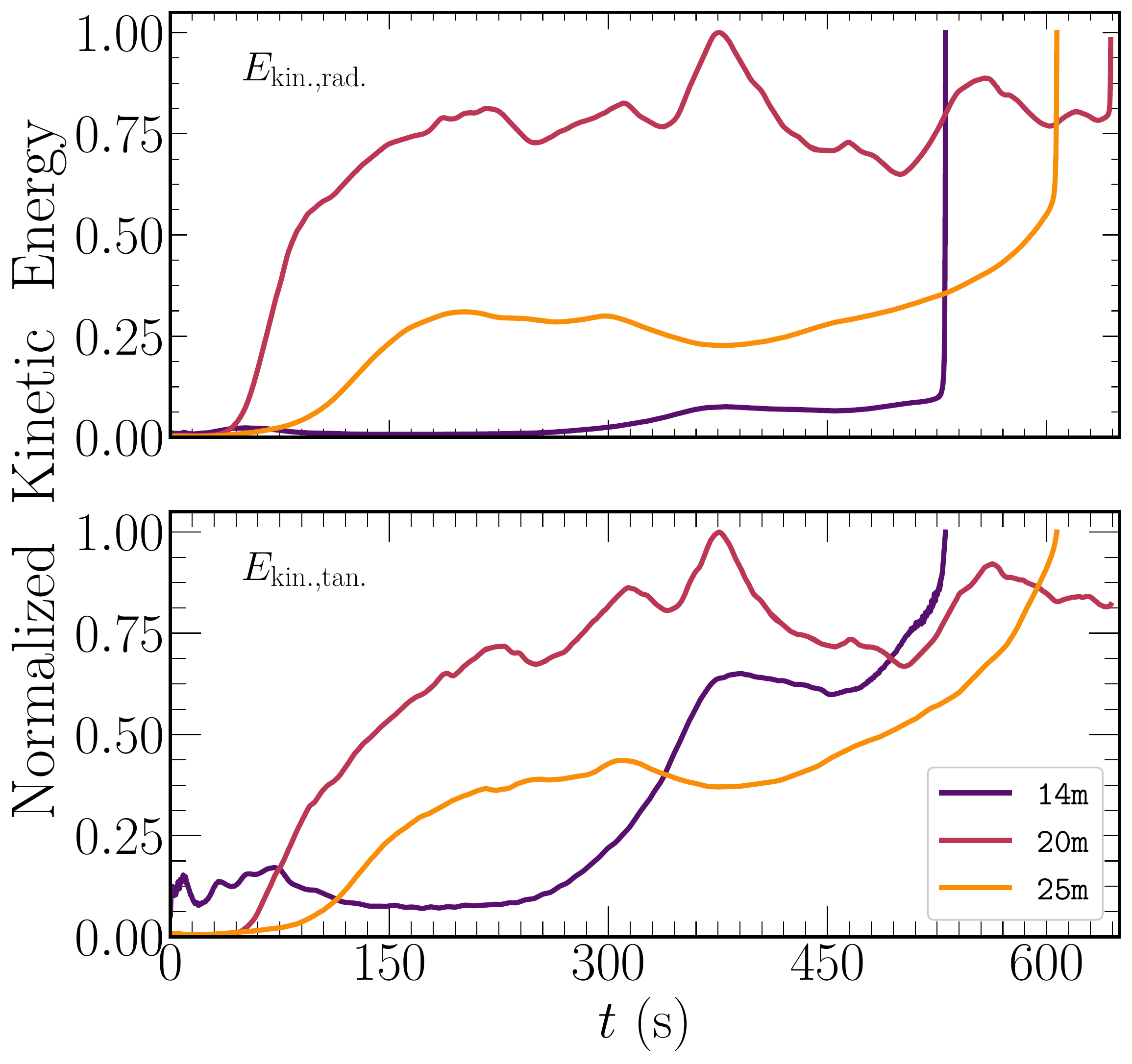}}
        \end{subfigure}
        \caption{Time evolution of the radial and tangential ($\theta + \phi$) kinetic energy throughout the simulation. For comparison
        between the different progenitor models, we normalize each simulation to the peak kinetic energy in the simulation. 
        }\label{fig:3d_e_kin}
\end{figure}

In Figure~\ref{fig:1d_isos} we show the convective velocity (top), oxygen-16 mass fraction 
(middle), and silicon-28 mass fraction (bottom) 
for the 1D \texttt{MESA} models at the time of mapping into \texttt{FLASH}. The 14 \msun and 20 \msun models, 
similar to the 15 \msun model from FC20 have a rather narrow Si-shell region. However, in the 
25 \msun model we see a Si-shell region that spans from $m \approx$ 1.9-2.4\msun. 
The 14 \msun and 15 \msun model have a similar O-shell region width and location. The 
20 \msun model has a much wider O-shell region that extends from $m \approx$ 1.4-4.0 \msun. 
The 25 \msun model also has a larger O-shell extending from $m \approx$ 2.4-5.7 \msun. 

The convection speeds as predicted by \MESA and mixing length theory (MLT) are largest 
in the O-shell region for the 20 \msun model with $v_{\rm{conv.}}\approx$ 120 km s$^{-1}$. The 
25 \msun model shows the second largest speeds with $v_{\rm{conv.}}\approx$ 70 km s$^{-1}$. 
The 14 \msun and 15 \msun model show similar, slower speeds of $v_{\rm{conv.}}\approx$ 40 km s$^{-1}$.
In all of the 1D models (except for the 25 \msun model), convection in the Si-shell region show 
speeds of $v_{\rm{conv.}}\lesssim$ 50 km s$^{-1}$. The 25 \msun model shows slower convective speeds 
of $v_{\rm{conv.}}\lesssim$ 35 km s$^{-1}$ at the time of mapping.

\section{3D Evolution To Iron Core-Collapse In Multiple Progenitors} 
\label{sec:results_all}
We evolve a total of four $4\pi$ 3D hydrodynamical massive star models for approximately
the final 10 minutes up to and including gravitational instability and iron core collapse. Three models 
with initial masses of 14 \msun, 20 \msun, and 25 \msun are evolved with velocity fields initialized using the methods 
and parameters described in \S~\ref{sec:methods}. To assess the impact of the initial velocity 
field topology on our results we also evolve an additional 20 \msun model using different spherical harmonic 
indices but otherwise the same parameters. 

Each simulation
is evolved up to the simulation time corresponding to iron core collapse as determined by \MESA. This
time corresponds with that at which the peak infall velocity within the iron core exceeds 1000 km s$^{-1}$. In the following
sections, we explore various aspects of our models including the general properties of three of our models, 
characterization of their convective shells, and comparisons of the angle-averaged profiles of the 3D 
simulations to the 1D results from \MESA. We explore the impact of the initial perturbations by 
comparing our two 3D 20 \msun models in \S~\ref{sec:results_20m}.

\begin{figure}[!t]
         \centering  
        \begin{subfigure}{
                \includegraphics[width=0.47\textwidth]{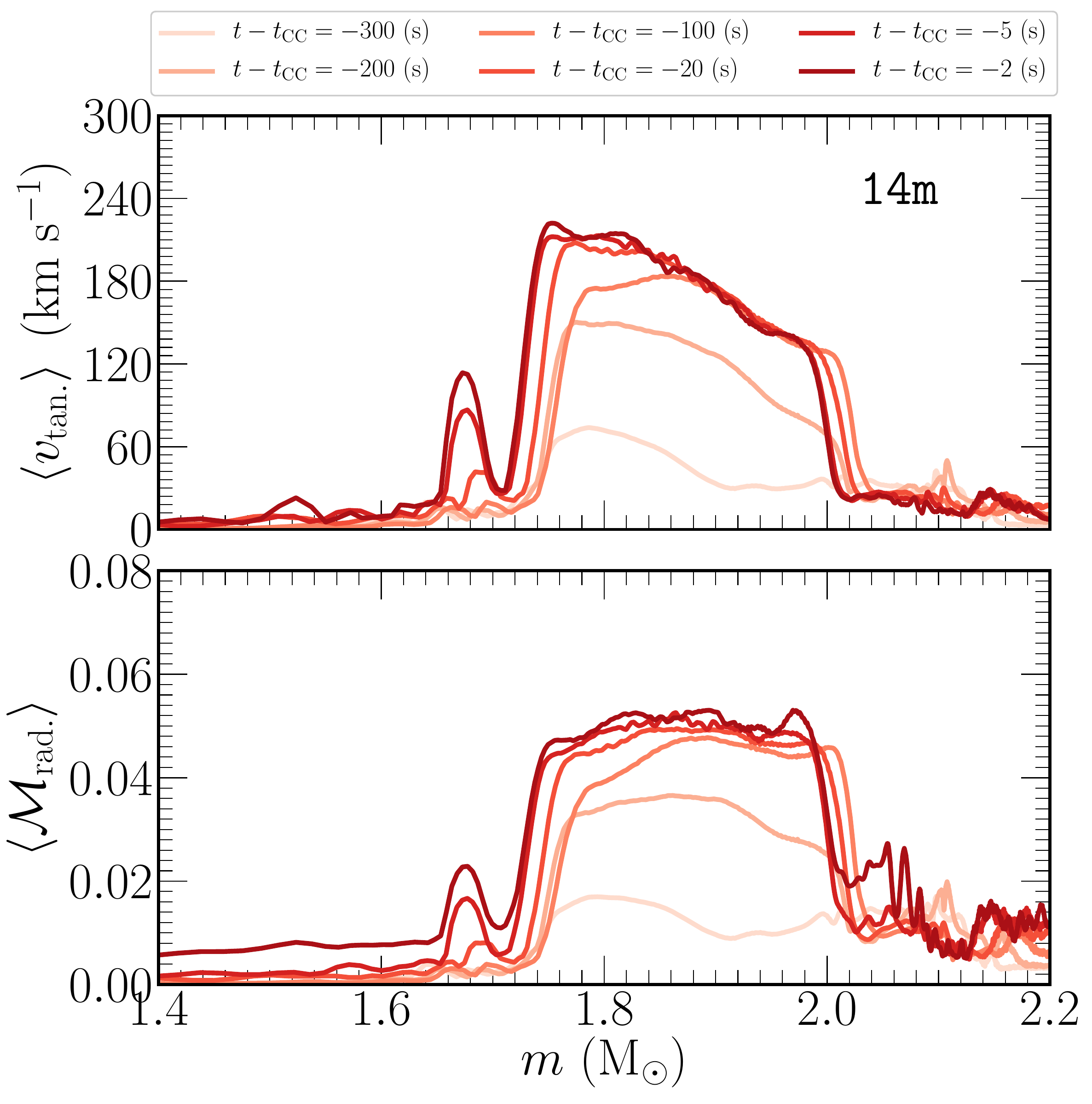}}
        \end{subfigure}
        \caption{Angle-average profiles of the tangential ($\theta + \phi$) velocity (top) and the radial Mach number (bottom) at six 
        different times for the 14 \msun model.
        }\label{fig:3d_14m_conv_props}
\end{figure}

\subsection{Global Properties}
\label{sec:all}

In Figure~\ref{fig:3d_e_kin} we show the time evolution of the radial and tangential components of the kinetic 
energy for the three 3D models of varied ZAMS mass. 
After an initial transient phase, the 14 \msun 
model reaches a quasi-steady state represented by a near constant value of 20\% of the local maximum at 
early times for the 
total tangential kinetic energy ($\approx 1 \times 10^{46}$ erg) until $t\approx$ 300 s.
In Figure~\ref{fig:3d_14m_conv_props} we show the angle-average tangential velocity and radial 
Mach number profiles for the 14-\msun model at six different times.  
Beyond 300 s (or $\approx 231$ s prior to core-collapse), the convective speeds in the 
O-shell region increase until a new local maxima in the kinetic energy is found near $t\approx$ 375 s.
At time $t-t_{\rm{CC}}=-100$ s ($\approx$ 430 s simulation time), the peak tangential velocity speeds are approximately 
160 km s$^{-1}$ throughout the shell with a corresponding
total kinetic energy of $E_{\rm{kin.}}\approx1\times10^{47}$ erg. Assuming a radius for the convective O-shell region of $r_{\rm{O}}\approx 7500$ km we can compute an approximate convective turnover time of 
$\tau_{\rm{conv.,O}}\approx 2 r_{\rm{O}} / v_{\rm{tan.,O}} \approx$ 94 s. The model undergoes
undergoes a few convective turnovers in the O-shell within this new state before core contraction 
continues to accelerate, increasing
the convective speeds further and pushing the model out of equilibrium as collapse ensues. 

For the 14-\msun model, we observe a radial Mach number (bottom panel of Figure~\ref{fig:3d_14m_conv_props}, 
where $\left< ... \right>$ denotes the angle-average of a particular quantity) of $\mathcal{M}_{\rm{rad.}}\approx0.05$ in the 
O-shell. In the Si-shell region the radial Mach number is
$\mathcal{M}_{\rm{rad.}}\lesssim0.01$ until the five seconds when a peak value of $\mathcal{M}_{\rm{rad.}}\approx0.02$
is reached. 
FC20 find a similar value in the O-shell of their 15-\msun model but a larger value in the
Si-shell region than observed here. Both of these models find radial Mach numbers that are about 50$\%$ 
than observed in the O-shell region of the 3D 18 \msun model of \citet{muller_2016_aa}.

\begin{figure}[!t]
         \centering  
        \begin{subfigure}{
                \includegraphics[width=0.47\textwidth]{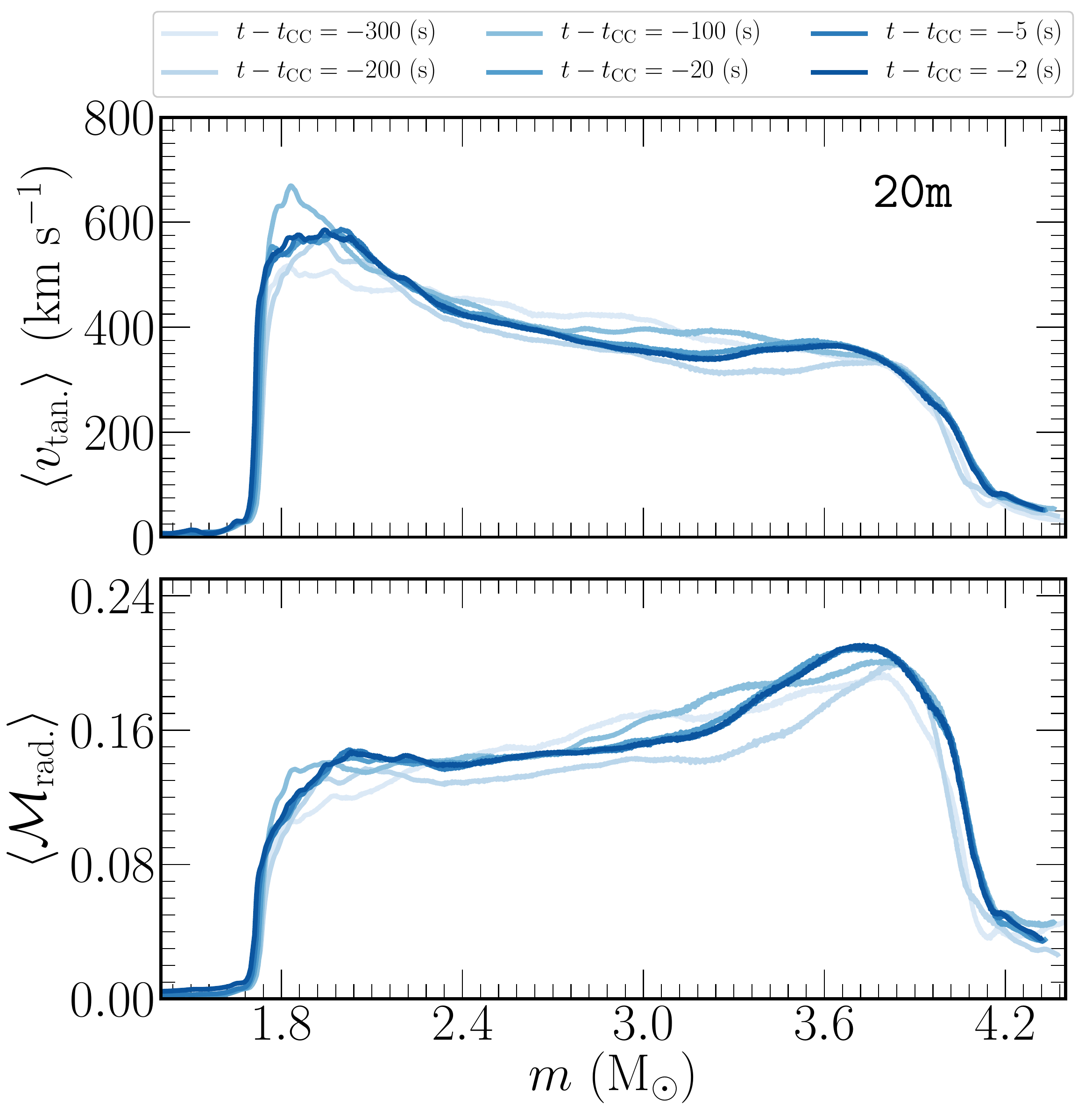}}
        \end{subfigure}
        \caption{Same as in Figure~\ref{fig:3d_14m_conv_props} but for the 20 \msun model.
        }\label{fig:3d_20m_conv_props}
\end{figure}

The 20-\msun model follows a qualitatively different evolution. For the first $\approx$ 200 s of the simulation, the 
radial and tangential components of the kinetic energy show a continued increase until reaching a saturation 
point. At this time, the model maintains a total kinetic energy value of $E_{\rm{kin.}}\approx6\times10^{48}$ erg. 
In Figure~\ref{fig:3d_20m_conv_props} we show convective velocity and Mach number profile for the 20 \msun model. 
As times as early as $t-t_{\rm{CC}}=-300$ ($t=\approx343$ s for the 20-\msun model), we observe tangential velocity 
speeds of $v_{\rm{tan.}}\approx$ 450 km s$^{-1}$ within the O-shell region. 
Using the O-shell radius, we obtain a convective turnover time of $\tau_{\rm{conv.,O}} \approx$ 172 s.
This suggests that our 20-\msun model completed approximately three full convective turnovers in the O-shell region. 
In this simulation, the Si-shell region is thin and experiences only very weak convection. Our discussion will be limited 
to the O-shell for this model. However, we do find significant mixing at the C/O shell interface in the 
20 \msun model. The properties of the 20-\msun model will be expanded upon and compared to the additional 20-\msun model with different
velocity field perturbations in \S~\ref{sec:results_20m}.

The 20-\msun model shows radial Mach values that are much larger than those of our 14-\msun model and the models of FC20. 
In Figure~\ref{fig:3d_20m_conv_props} (bottom panel), we find a value of $\mathcal{M}_{\rm{rad.}}\approx0.15$ at the base of the
convective O-shell ($m\approx1.8$ \msun) and $\mathcal{M}_{\rm{rad.}}\approx0.21$ at the edge fo the O-shell / C/O shell interface.
These values are relatively constant in profile shape and magnitude for the final 300 seconds prior to collapse unlike
our 14 \msun model which grow in magnitude up to the final seconds prior to collapse. 

\begin{figure}[!t]
         \centering  
        \begin{subfigure}{
                \includegraphics[width=0.47\textwidth]{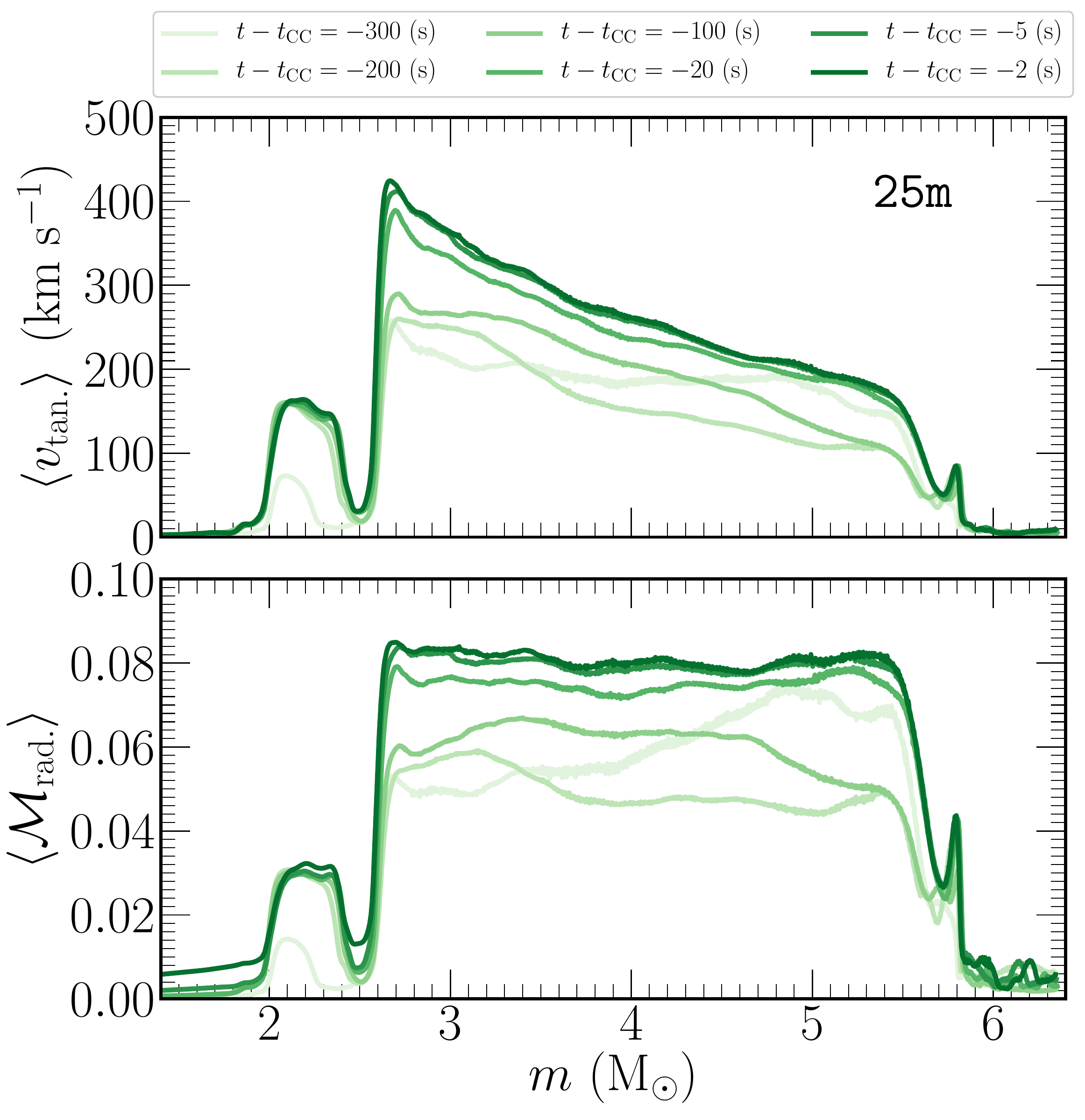}}
        \end{subfigure}
        \caption{Same as in Figure~\ref{fig:3d_14m_conv_props} but for the 25 \msun model.
        }\label{fig:3d_25m_conv_props}
\end{figure}

\begin{figure*}[!htb]
         \centering  
        \begin{subfigure}{
                \includegraphics[width=0.95\textwidth]{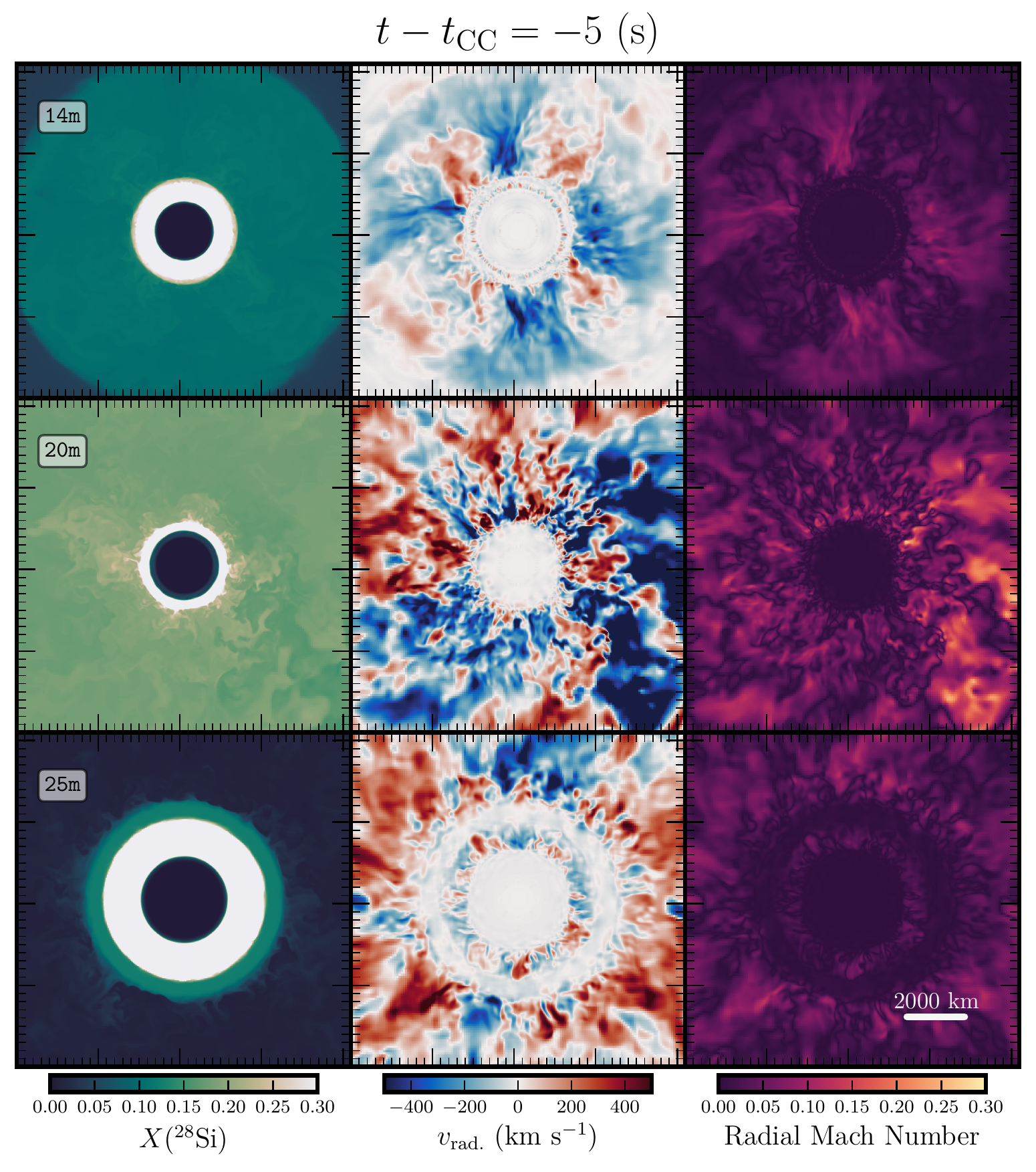}}
        \end{subfigure}
        \caption{Slice plot for the specific $^{28}$Si mass fraction (left column),
the radial velocity (middle column), and the radial mach number (right column) in the $x-y$ plane at a time 
approximately 5 seconds before iron core-collapse for three of our 3D models. 
        }\label{fig:3d_all_slice}
\end{figure*}

\begin{figure*}[!htb]
         \centering  
        \begin{subfigure}{
                \includegraphics[width=0.48\textwidth]{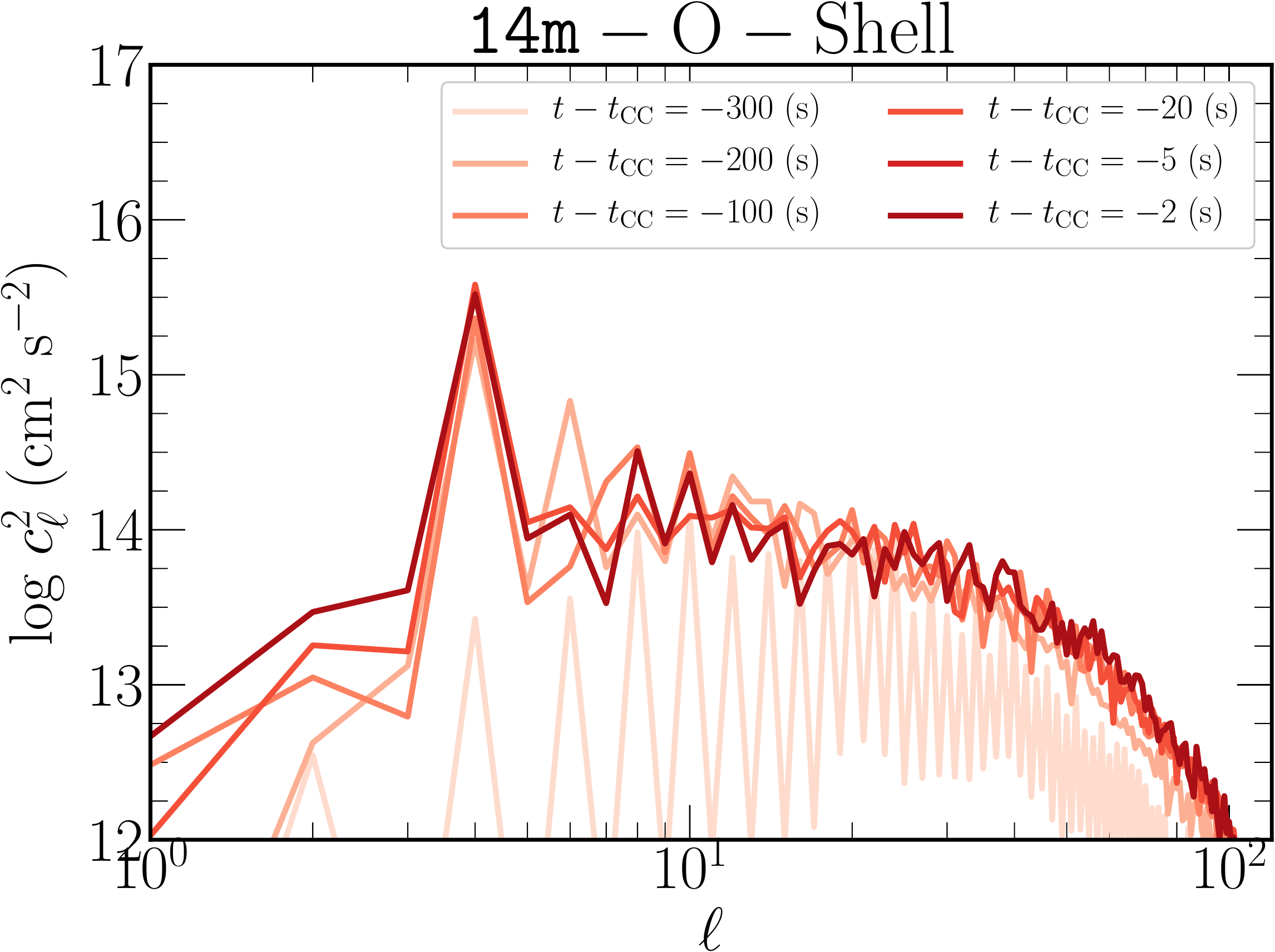}}
        \end{subfigure}
                \begin{subfigure}{
                \includegraphics[width=0.48\textwidth]{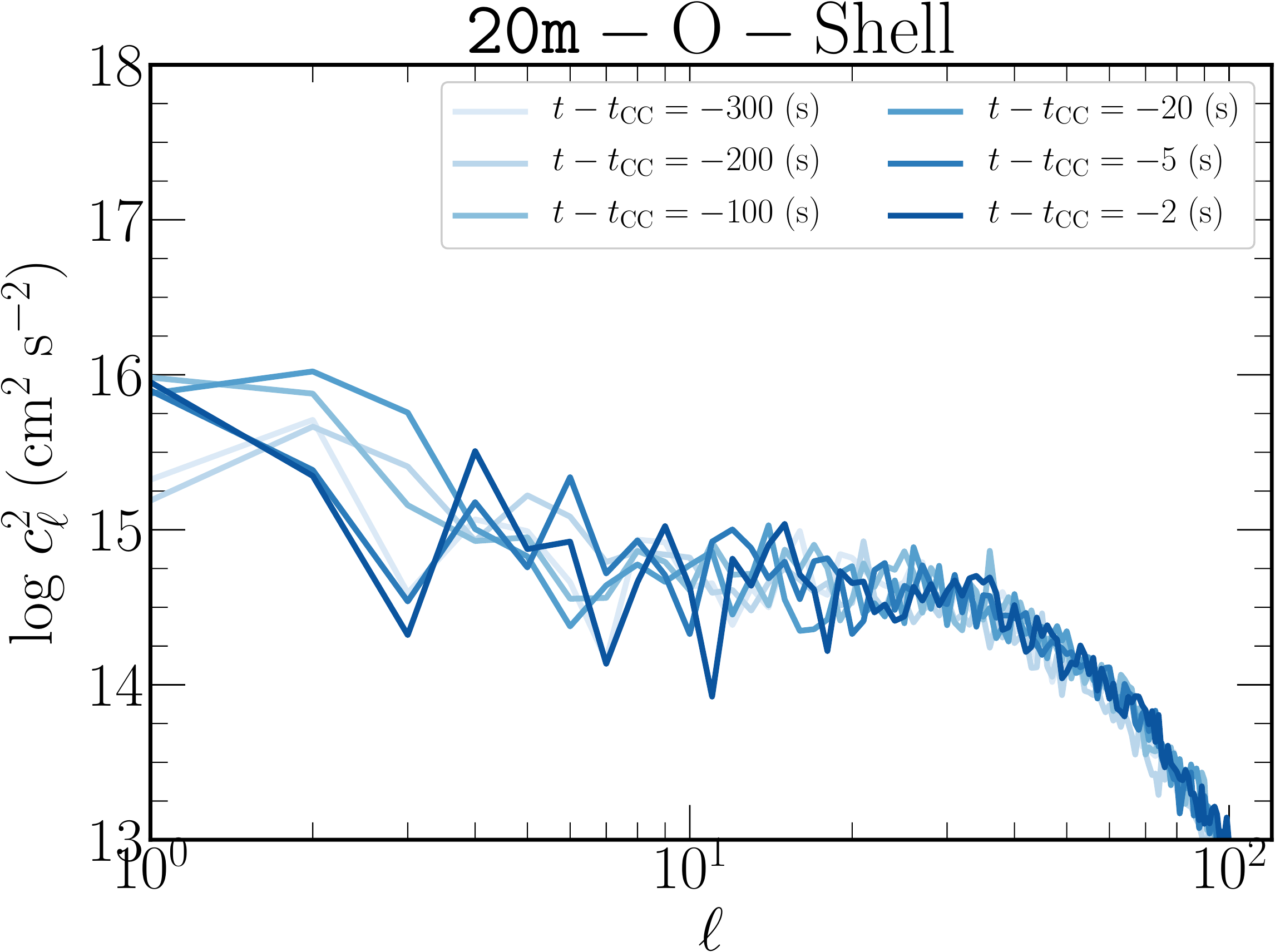}}
        \end{subfigure}
                \begin{subfigure}{
                \includegraphics[width=0.48\textwidth]{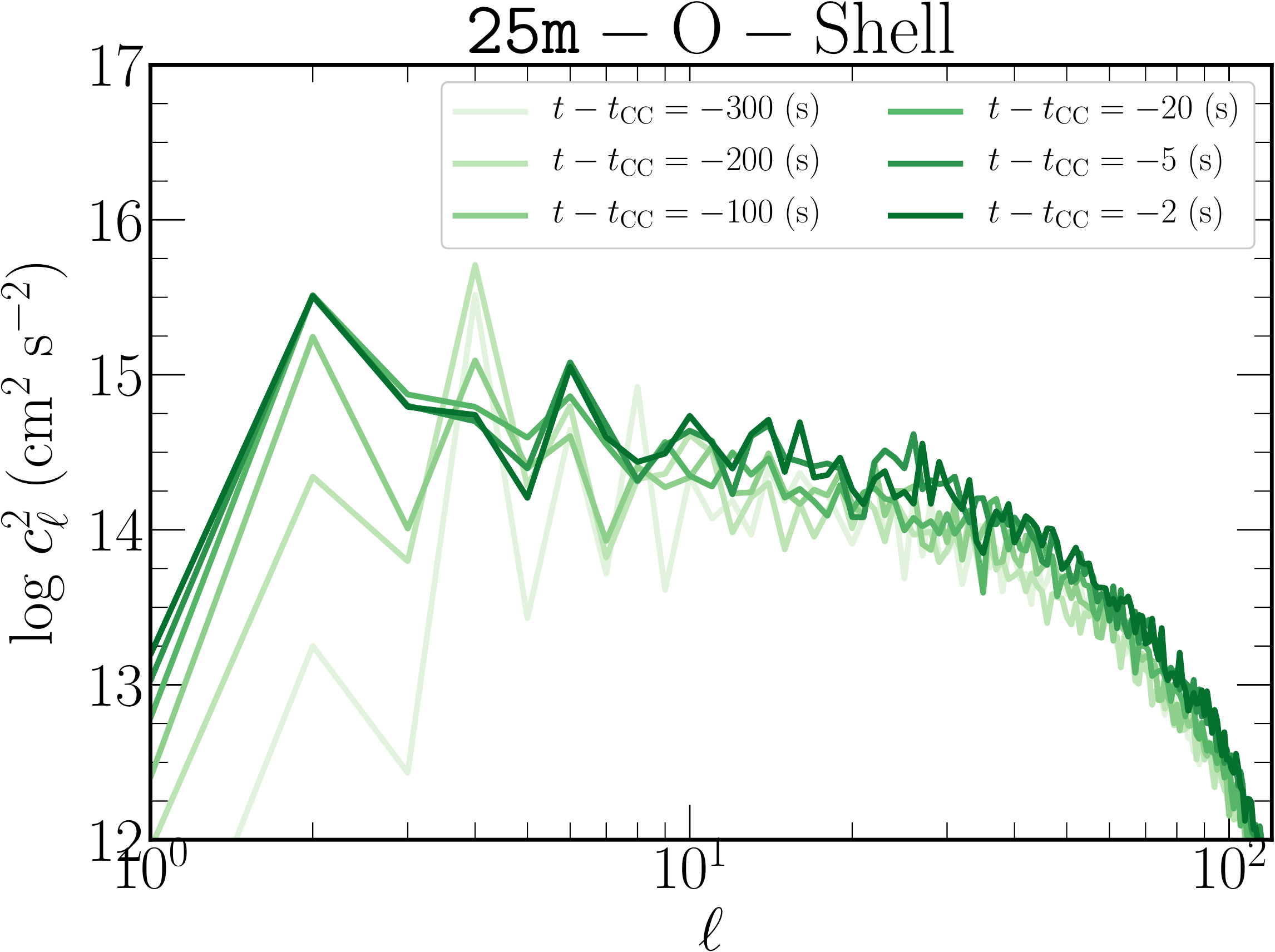}}
        \end{subfigure}
        \caption{Power spectrum of the radial velocity decomposition in the convective O-shell region for three of the 3D models.
        }\label{fig:3d_power_spectra}
\end{figure*}

Our 25 \msun model follows an evolutionary path that is different from the other two models. The model 
reaches about 30\% of the peak radial kinetic energy at a time of $t\approx$ 150 s. The kinetic energy fluctuates 
only slightly once reaching this saturated value of total kinetic energy corresponding to a value of 
$E_{\rm{kin.}}\approx 1\times 10^{48}$ erg. In Figure~~\ref{fig:3d_25m_conv_props} we show convective profiles now 
for the 25 \msun model. We can estimate a convective turnover time for the O-shell 
utilizing the tangential velocity speeds of $v_{\rm{tan.}}\approx$ 240 km s$^{-1}$ observed from $t\geq$ 150.
The speeds observed in the convective O-shell region lead to a turnover time of $\tau_{\rm{conv.,O}}\approx $ 320 s.
The turnover time in the O-shell region suggests that our model completed one full turnovers from $t\approx$ 150 to 500 s.
At $t\approx$ 500 s and beyond, the acceleration of the contracting iron core leads to a gradual increase of both of the
components of the kinetic energy relative to their max values at collapse. In this model, we find a larger 
Si-shell region where convection becomes efficient and relevant to the dynamics of the model near collapse.
At $t\approx$ 300 s, the width of the Si-shell expands and the convective velocity speeds begin to increase. The 
speeds in this shell region saturate at a value of $v_{\rm{tan.}}\approx$ 160 km s$^{-1}$ at $t\approx$ 400 s, maintaining
this value until core-collapse. Within the Si-shell region, we find a convective turnover time of 
$\tau_{\rm{conv.,Si}}\approx $ 6 s suggesting that from $t\approx$ 400 s to core-collapse, our 25 \msun
captures 34 convective turnovers. The 25 \msun model shows radial Mach numbers of $\mathcal{M}_{\rm{rad.}}\approx0.03$ 
in the Si-shell and $\mathcal{M}_{\rm{rad.}}\approx0.08$ in the O-shell region in the final 20 seconds prior to collapse.  

In Figure~\ref{fig:3d_all_slice} we show the a pseudocolor slice plot for the specific $^{28}$Si mass fraction (left column),
the radial velocity (middle column), and the radial mach number (right column) in the $x-y$ plane at a time 
approximately 5 seconds before iron core-collapse. The differences between the three simulations are shown 
here mainly in the relative size of the O-shell regions, convective flow properties, and mixing within the different shell
regions.

\subsection{Power Spectrum Of Convective Shells}
To further quantify the convective properties in our 3D models, we decompose the 
perturbed radial velocity field into spherical harmonics for the O-shell region (and also the Si-shell region for the 25 \msun model).
Similar to FC20, the total power for a given spherical harmonic order, $\ell$, as 
\begin{equation}
c^{2}_{\ell} = \sum\limits^{\ell}_{m=-\ell} \left | \int Y^{m}_{\ell} (\theta,\phi) 
v_{\rm{rad.}}^{\prime}(r_{\rm{Shell}},\theta,\phi) d\Omega \right | ^{2}~,
\end{equation}
where $v_{\rm{rad.}}^{\prime}$ = $v_{\rm{rad.}}$ - $\tilde{v}_{\rm{rad.}}$, with $\tilde{v}_{\rm{rad.}}$ corresponding to the mean background radial velocity speed 
at the chosen shell radius \citep{sht_2013_aa}. In the 14 \msun, 20 \msun, and 25 \msun models the O-shell regions 
are evaluated at $r_{\rm{O}}=$5000 km, $r_{\rm{O}}=$5000 km, and $r_{\rm{O}}=$10 000 km, respectively. In 
the 25 \msun model, the Si-shell region is evaluated at $r_{\rm{Si}}=$3900 km.
The expected dominant mode can be also approximated using 
\begin{equation}
\ell  = \frac{\pi}{2} \left ( \frac{r_{+} + r_{-}}{r_{+} - r_{-}} \right )~,
\label{eq:ell}
\end{equation}
where $r_{+}$ and $r_{-}$ are the upper and lower shell radii, respectively \citep{foglizzo_2006_aa}. 
In Figure~\ref{fig:3d_power_spectra} we show the resulting O-shell power spectra for three of our 3D models. 

The 14 \msun 
model shows a relatively constant power spectrum during the last 100 s prior to collapse, the spectra is peaked 
at a spherical harmonic index of $\ell=4$. Before this, at 300 seconds prior to collapse, convective is relatively 
underdeveloped and power is significantly less across scales. At $t-t_{\rm{CC}}=-200$ s, the spectrum shows a 
peak at $\ell=7$, the driving scale due to our initial perturbations. Energy is then transferred to larger scales at 
$\ell=4$ and remains there for the duration of the simulations. As the simulation approaches collapse, we observe 
a slight increase in 
power at larger scales $\ell=1-3$. The dominant mode predicted by using Equation~\ref{eq:ell}
is found to be $\ell\approx2.38$. The excess in power at $\ell=4$ is likely attributed to the Cartesian nature of our grid 
geometry and increased numerical viscosity near the grid axes.

\begin{figure}[!t]
\centering{\includegraphics[width=1.0\columnwidth]{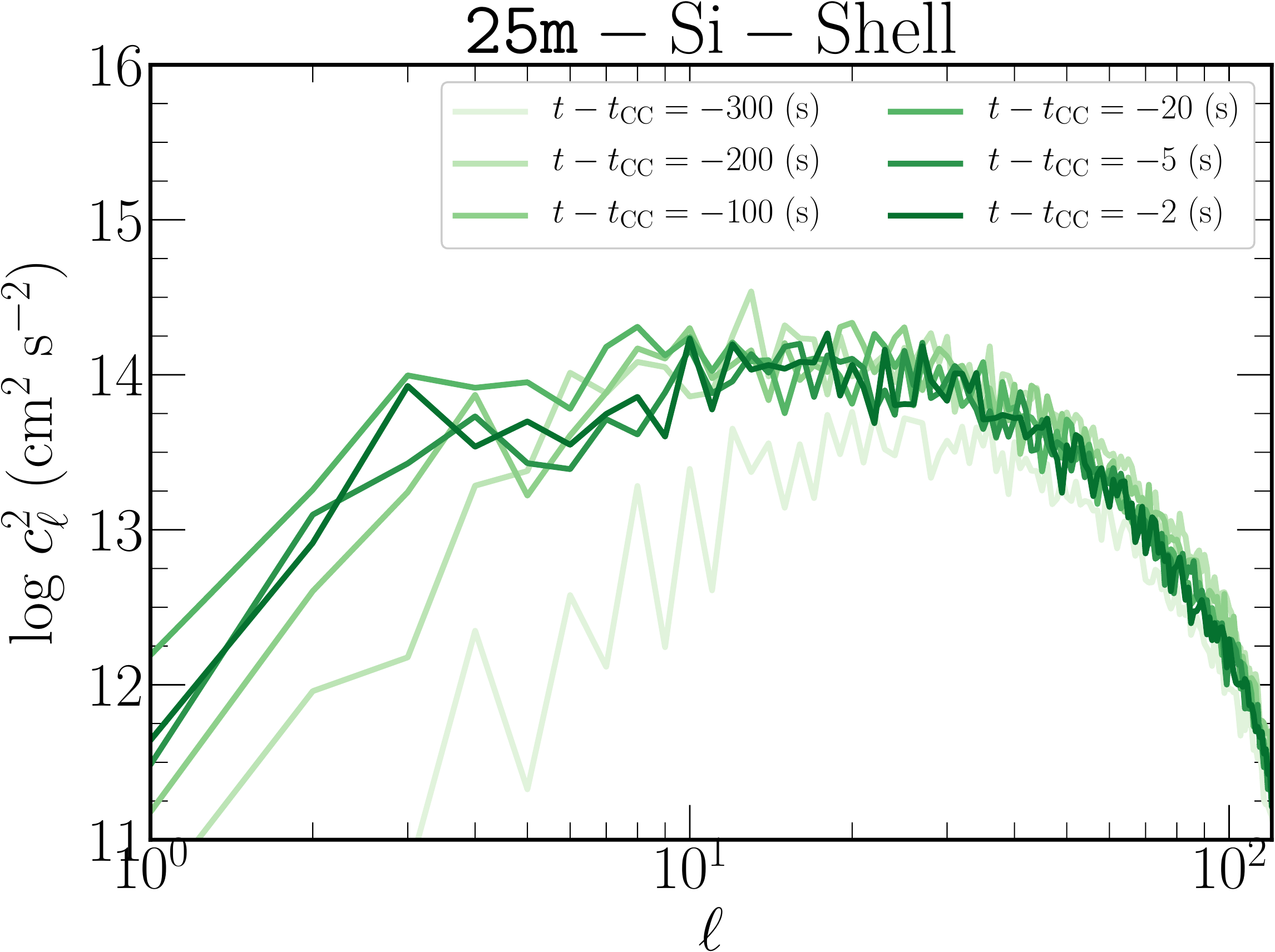}}
\caption{Same as in Figure~\ref{fig:3d_power_spectra} but for the 25 \msun Si-shell region. 
}\label{fig:3d_power_spectra_si}
\end{figure}

The power spectrum for the 20 \msun model (top right of Figure~\ref{fig:3d_power_spectra}) shows a significantly different
qualitative evolution towards collapse. At $t-t_{\rm{CC}}=-200$ s, the power is distributed across many scales with a characteristic
peak at $\ell=2$ suggesting the flow is dominated by a large scale quadrupole flow structure. This results also 
agrees with Equation~\ref{eq:ell} which predicts a dominant mode of $\ell\approx1.82$. A similar dominant mode was observed in the 
18 \msun model of \citet{muller_2016_aa} in the final moments prior to collapse. At $t-t_{\rm{CC}}=-100$ s, energy in this
mode and the $\ell=1$ mode increase, with $\ell=1$ becoming the dominate mode. 
At later times closer to collapse, the 
peak at $\ell=2$ decreases in power with scales of $\ell=1,4$ increasing in power. In the final 5 seconds 
prior to collapse, the dominant mode is observed at $\ell=1$ with power in intermediate scales of $\ell=4-8$.

\begin{figure*}[!htb]
         \centering  
        \begin{subfigure}{
                \includegraphics[width=0.475\textwidth]{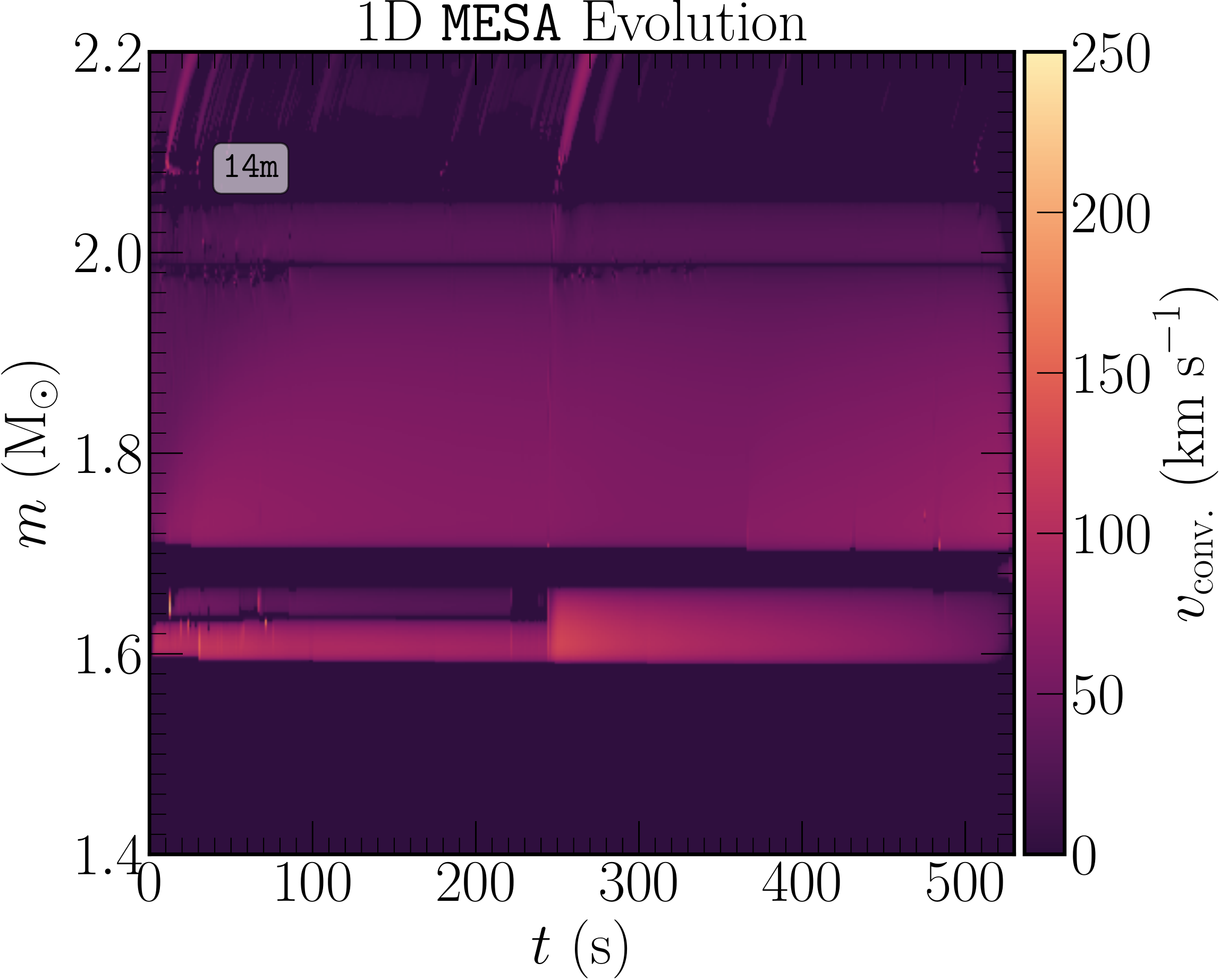}}
        \end{subfigure}
        \begin{subfigure}{
                \includegraphics[width=0.475\textwidth]{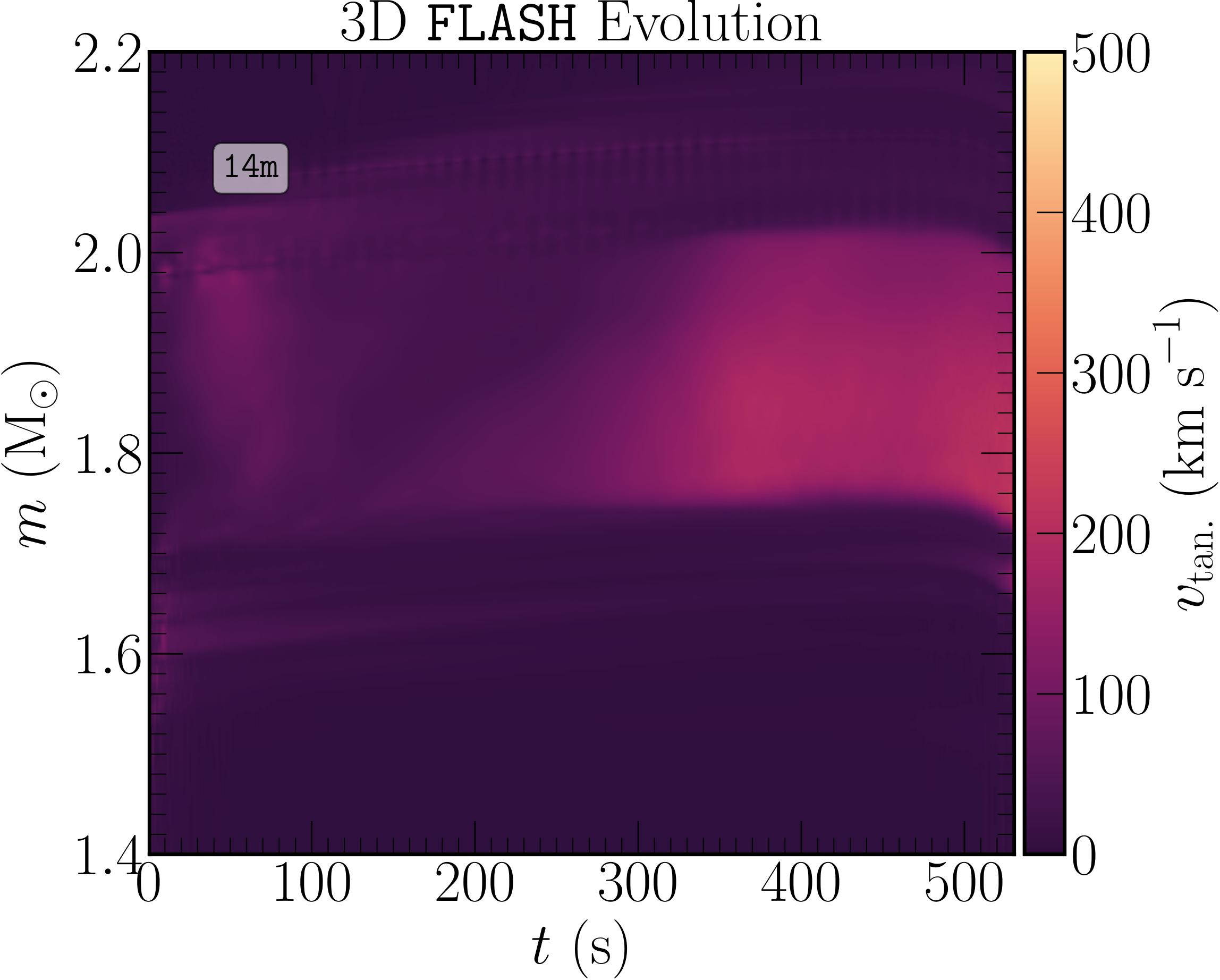}}
        \end{subfigure}
        \begin{subfigure}{
                \includegraphics[width=0.475\textwidth]{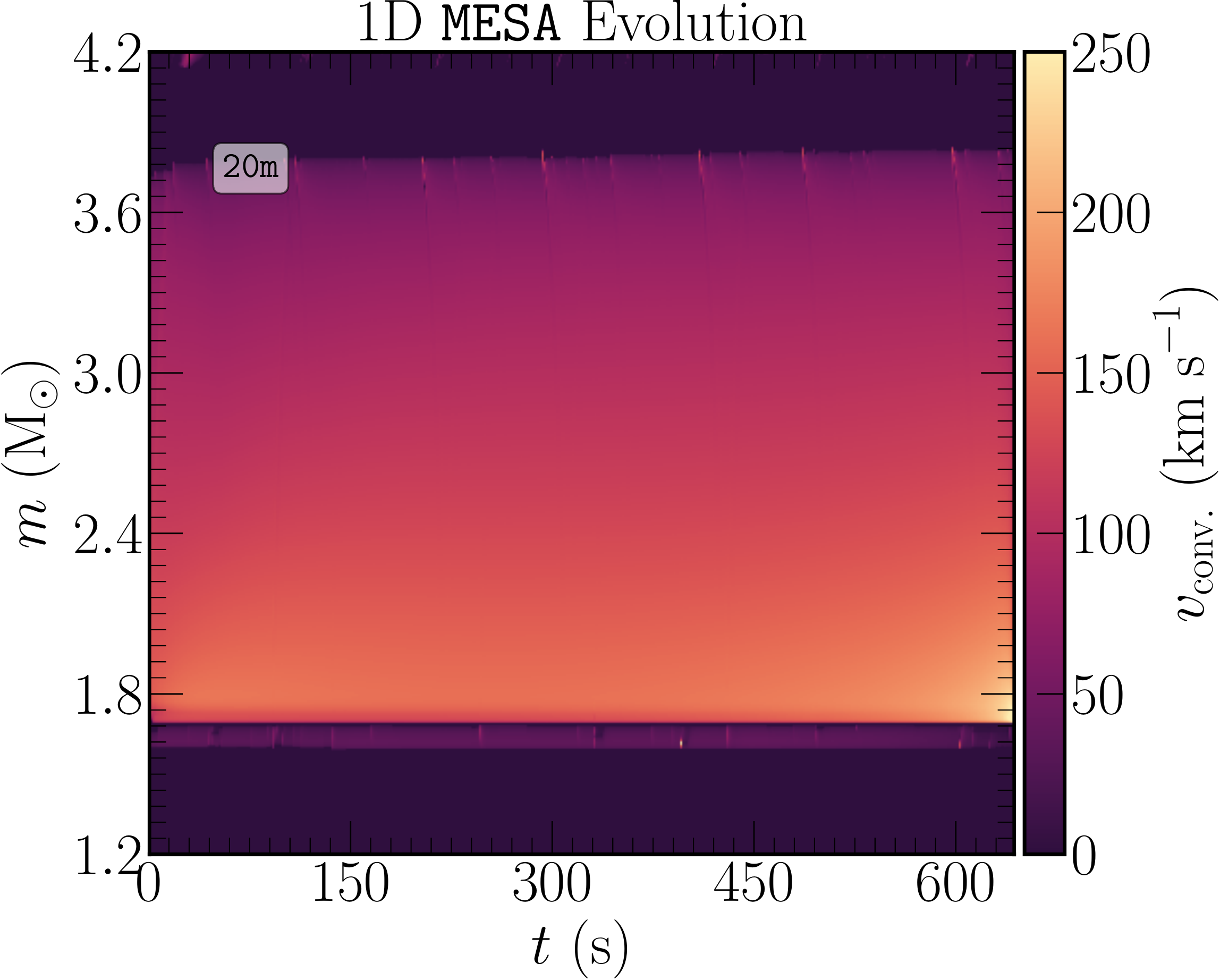}}
        \end{subfigure}
        \begin{subfigure}{
                \includegraphics[width=0.475\textwidth]{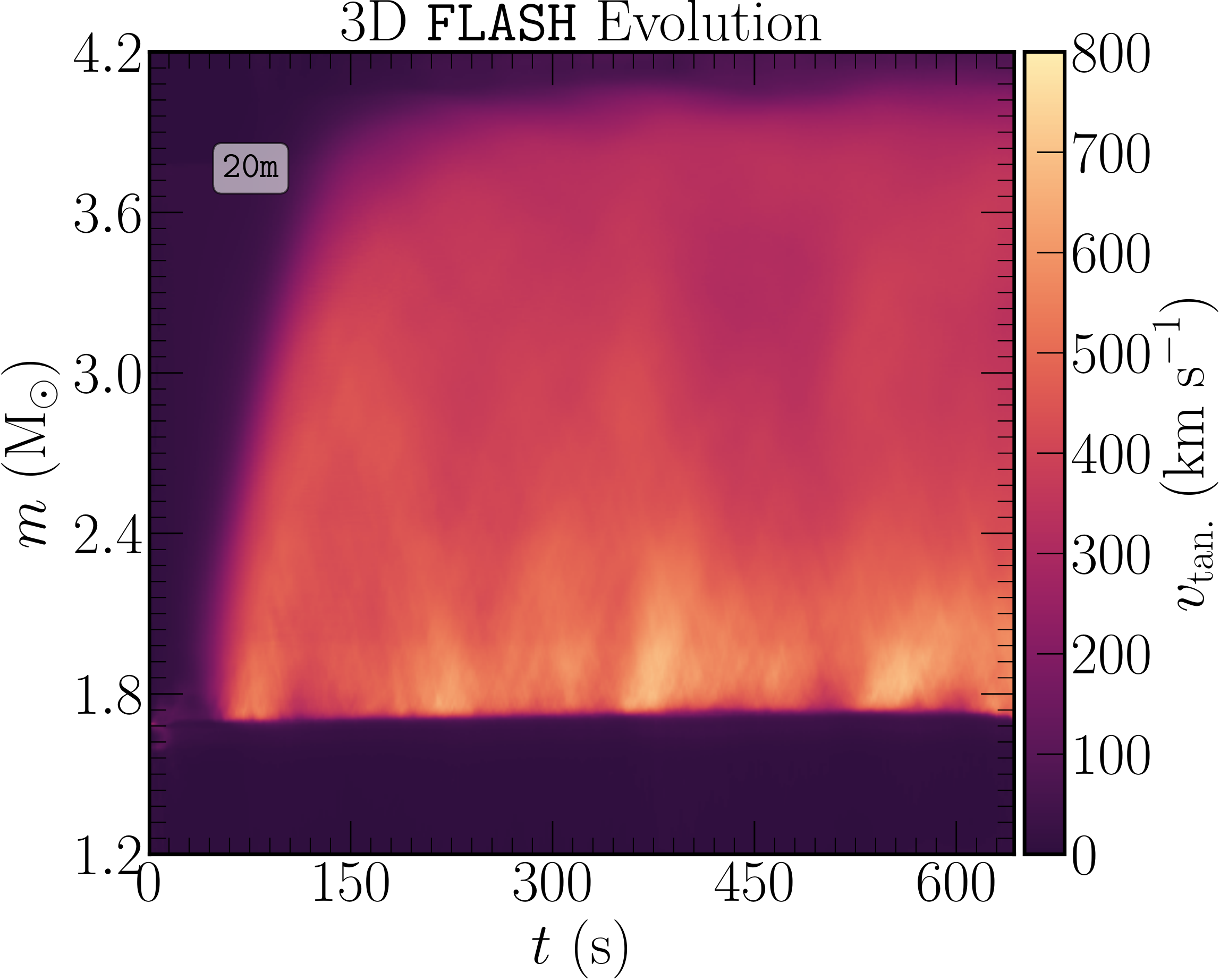}}
        \end{subfigure}
        \begin{subfigure}{
                \includegraphics[width=0.475\textwidth]{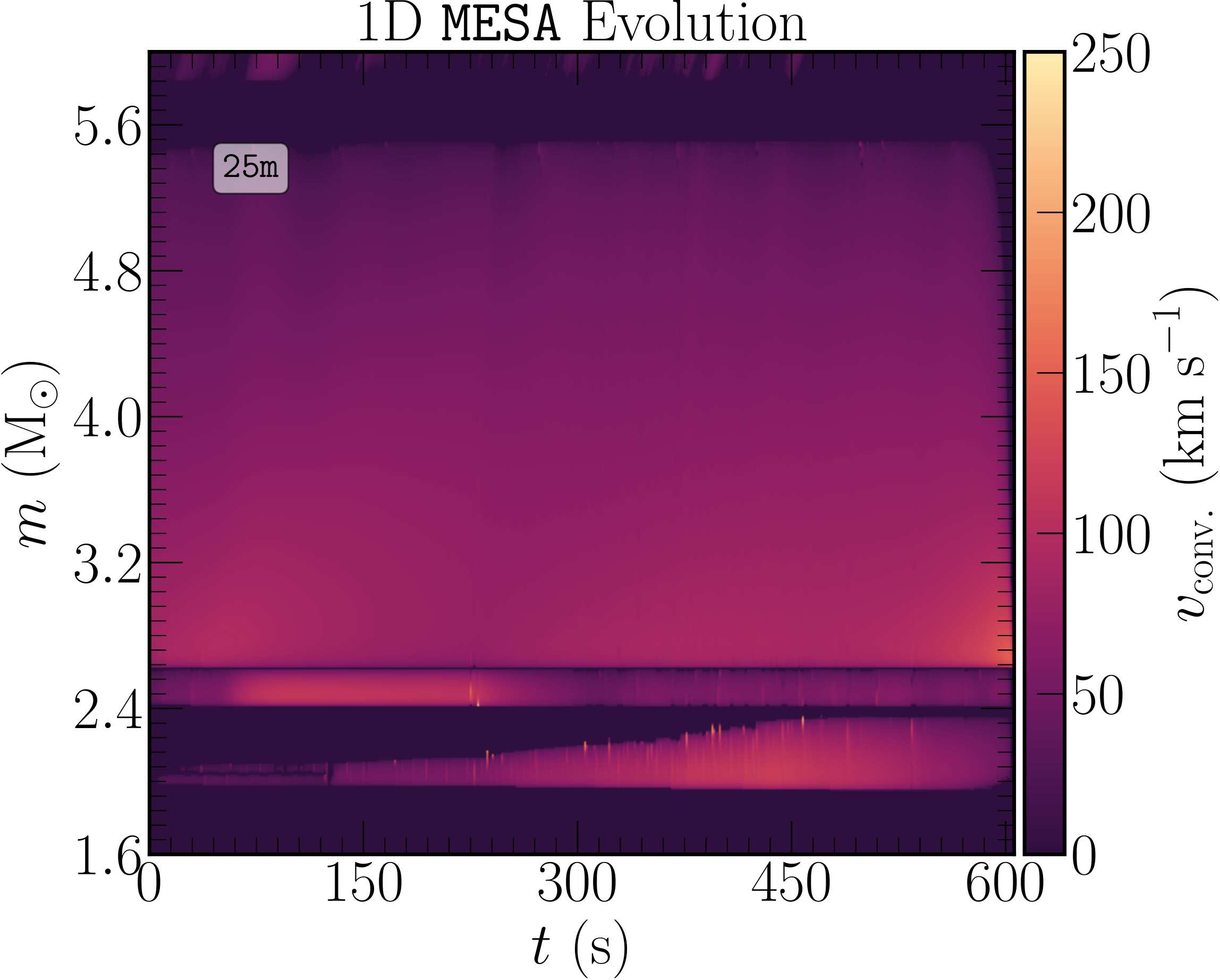}}
        \end{subfigure}       
        \begin{subfigure}{
                \includegraphics[width=0.475\textwidth]{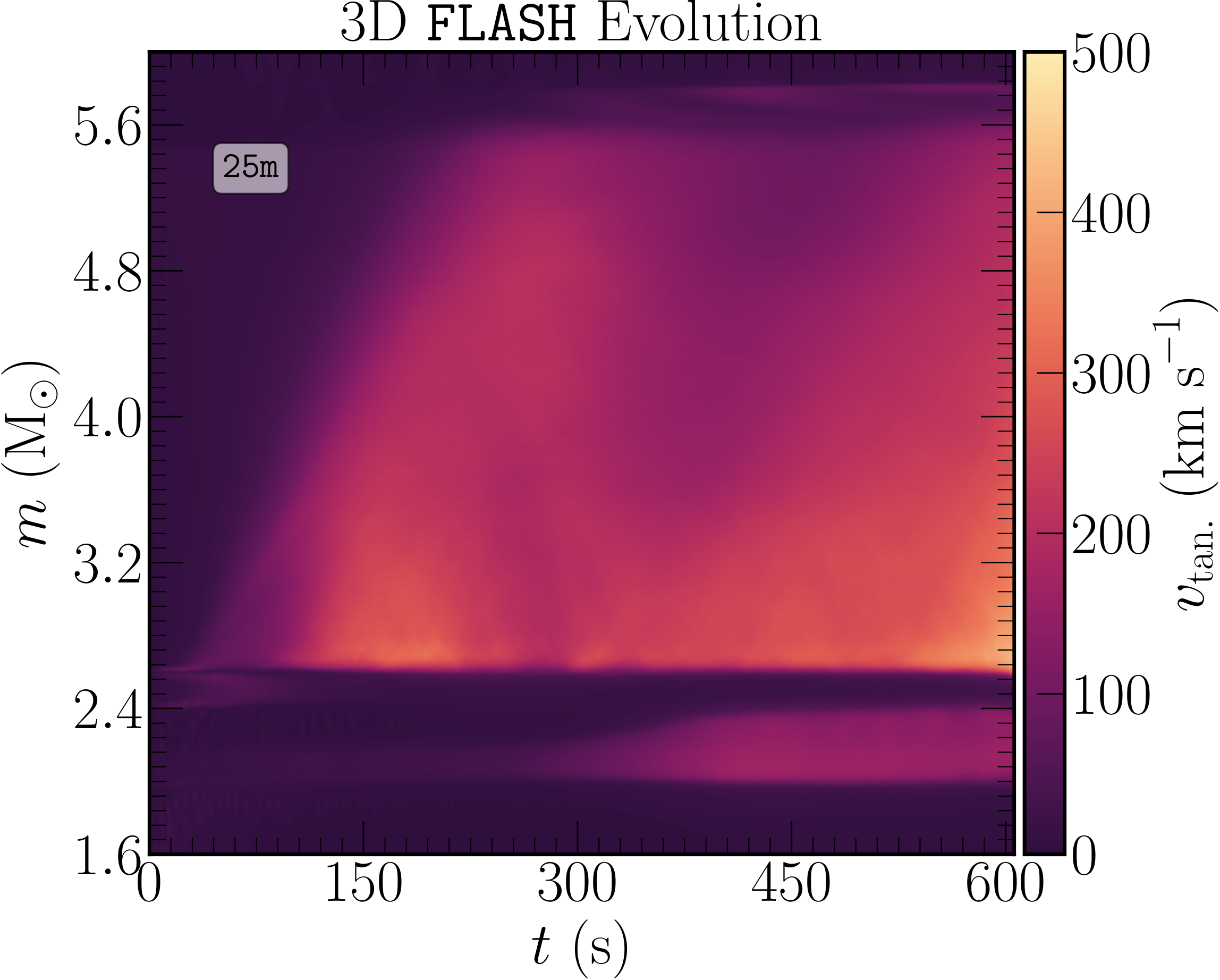}}
        \end{subfigure}   
        \caption{Pseudocolor heatmaps of the convective velocity profiles according to the 1D \MESA models (left column) compared 
        the angle-averaged profiles of the tangential velocity component for the three 3D \FLASH simulations (right column). In all case, 
        the scale for the 3D simulations is more than twice that of the \MESA models for our choice of mixing length parameter of $\alpha_{\rm{MLT}}=1.5$.
        }\label{fig:heatmaps_v_conv}
\end{figure*}

\begin{figure}[!t]
         \centering  
        \begin{subfigure}{
                \includegraphics[width=0.47\textwidth]{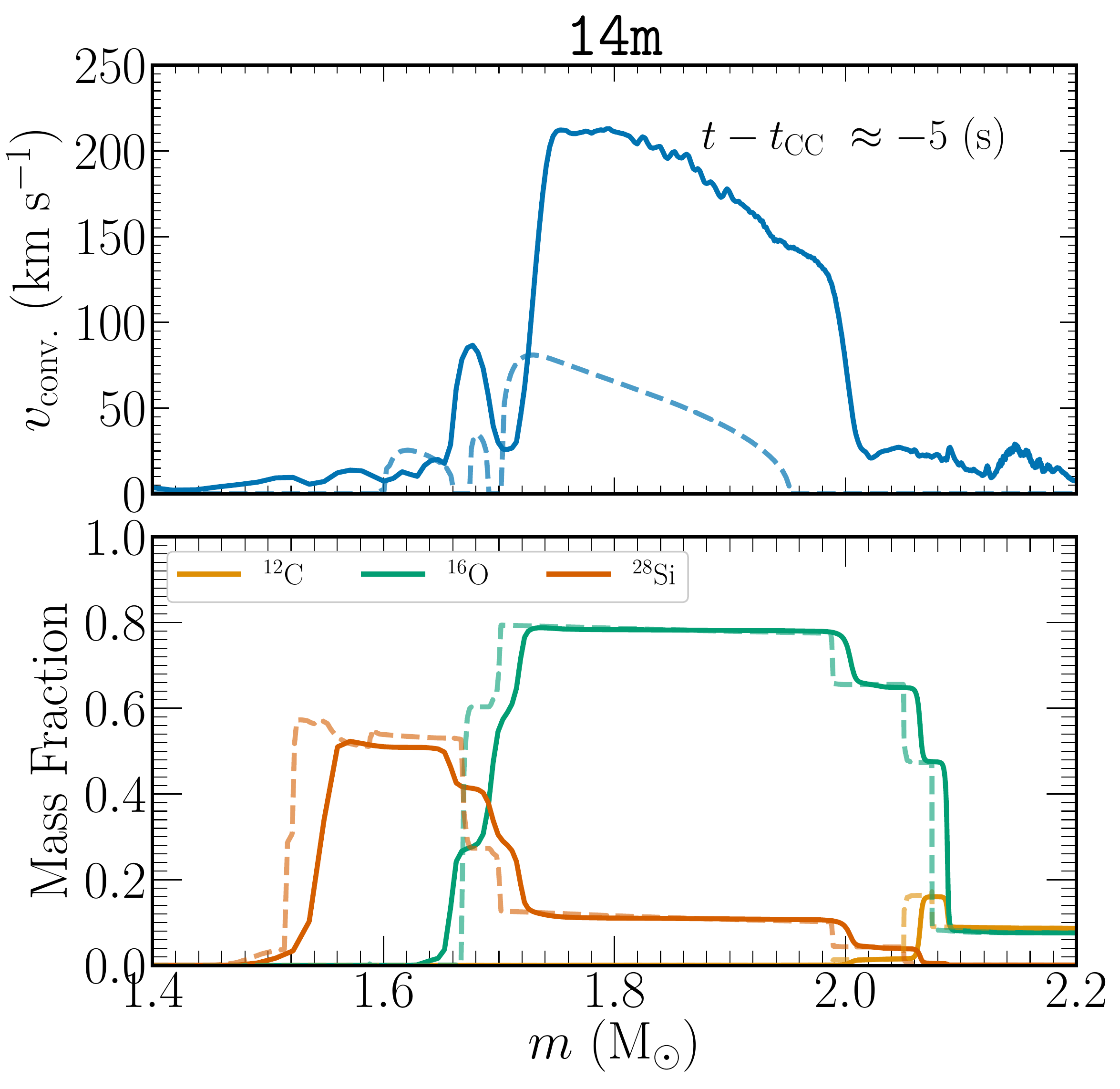}}
        \end{subfigure}
        \caption{Angle-averged tangential velocity (top) and mass fraction profiles for $^{12}$C, 
        $^{16}$O, and $^{28}$Si compared to the 1D profiles from \MESA at a time approximately 5 seconds
        prior to iron core-collapse. The dashed line in both plots corresponds to the \MESA profile and the solid line
        corresponds to the 3D \FLASH simulation.        
        }\label{fig:14m_compare_mesa}
\end{figure}

The 25 \msun simulation shows a qualitative evolution towards collapse in the convection shell that has traits of the 
14 \msun and 20 \msun models. At early times of $t-t_{\rm{CC}}=-300$ s and $t-t_{\rm{CC}}=-200$ s, an excess
of power is again observed at an $\ell=4$ mode suggesting grid aligned artifacts contributing to the power spectrum of the 
radial velocity field. However, at times beyond this, the 25 \msun model shows a reduction in power at this mode with a 
shift in the peak power contribution occurring at $\ell=2$. The predicted dominant mode is found to be $\ell\approx2.01$. 
The peak spherical harmonic mode remains $\ell=2$ for the final 100 s of the simulation with a slight contribution at the 
driving scale ($\ell=7$) observed in the final 5 seconds before core collapse. 

Figure~\ref{fig:3d_power_spectra_si} shows the power spectrum of the radial velocity field in the Si-shell 
region of the 25 \msun model at five different times. Overall, the Si-shell region shows less power across scales.
Convection begins to contribute to the power spectrum at times beyond 200 s prior to collapse. At this time, the 
convective Si-shell region has also expanded in radius to allow for a larger scale dominant mode to emerge. 
Using the approximate shell radii at $t-t_{\rm{CC}}=-100$ s, we find a predicted dominant mode of $\ell\approx10.09$.
The spectrum at this point is characterized by a broad range of power at intermediate scales from $\approx\ell=6-15$. 
At later times, we observe a slight increase in power at larger scales near $\ell=2-4$. Five seconds prior to collapse, 
the dominant modes are found to be found near spherical harmonic indices of $\ell=15-30$.

\subsection{Comparisons with 1D \MESA models}
An important aspect of our 3D will be their ability to inform 1D \MESA models of CCSN 
progenitors. It was shown in FC20, that the 3D simulations found angle averaged convective 
speeds that were on the order of \emph{four} times larger than predicted by \MESA. Larger non-radial convective velocity speeds 
in 3D CCSN progenitors than their 1D counterparts have implications that can lead to favorable conditions for explosion. 
Specifically, increased non-radial convective speeds can work to increase the mass in the 
gain region which can indirectly alter the effective neutrino heating rates and contribute to greater turbulent stresses in the gain layer \citep{couch_2013_aa,couch_2015_ab}.
Here, we compare angle-average profiles from our full 4$\pi$ 3D simulations to their 1D \MESA counterparts.

In Figure~\ref{fig:heatmaps_v_conv} we show the time evolution of the angle-averaged tangential velocity
component for each 3D model (right column) and the time evolution of the corresponding 1D \MESA 
convective velocity profiles predicted by MLT. 
The angle-average properties are weighted by the corresponding cell mass 
in each bin. For the profiles used to produce these heat maps we use $N=2048$ bins and use linear
interpolation to smooth the raw profiles. Qualitatively, most of the 3D models agree well with the \MESA predictions. 
The largest differences are found for the 14 \msun model where the convective region appears to 
expand and contract for the first few hundred seconds of the simulation at which point the tangential 
velocity speeds reach $v_{\rm{tan.}}\approx$ 200-300 km/s. The extent of the O-shell in the 20 \msun model
differs somewhat from the \MESA model due to mixing at the convective boundary layer between the O- and 
C-shell regions. This mixing will be examined further in \S~\ref{sec:results_20m}. The 25 \msun matches
 the \MESA model qualitatively well, where we see a slight expansion of the Si-shell region predicted by 
\MESA also shown in the 3D model at $t\approx350$ s. In the following subsections, we will explore 
each 3D model compared to its \MESA counterpart in more detail.

\begin{figure}[!t]
         \centering  
        \begin{subfigure}{
                \includegraphics[width=0.47\textwidth]{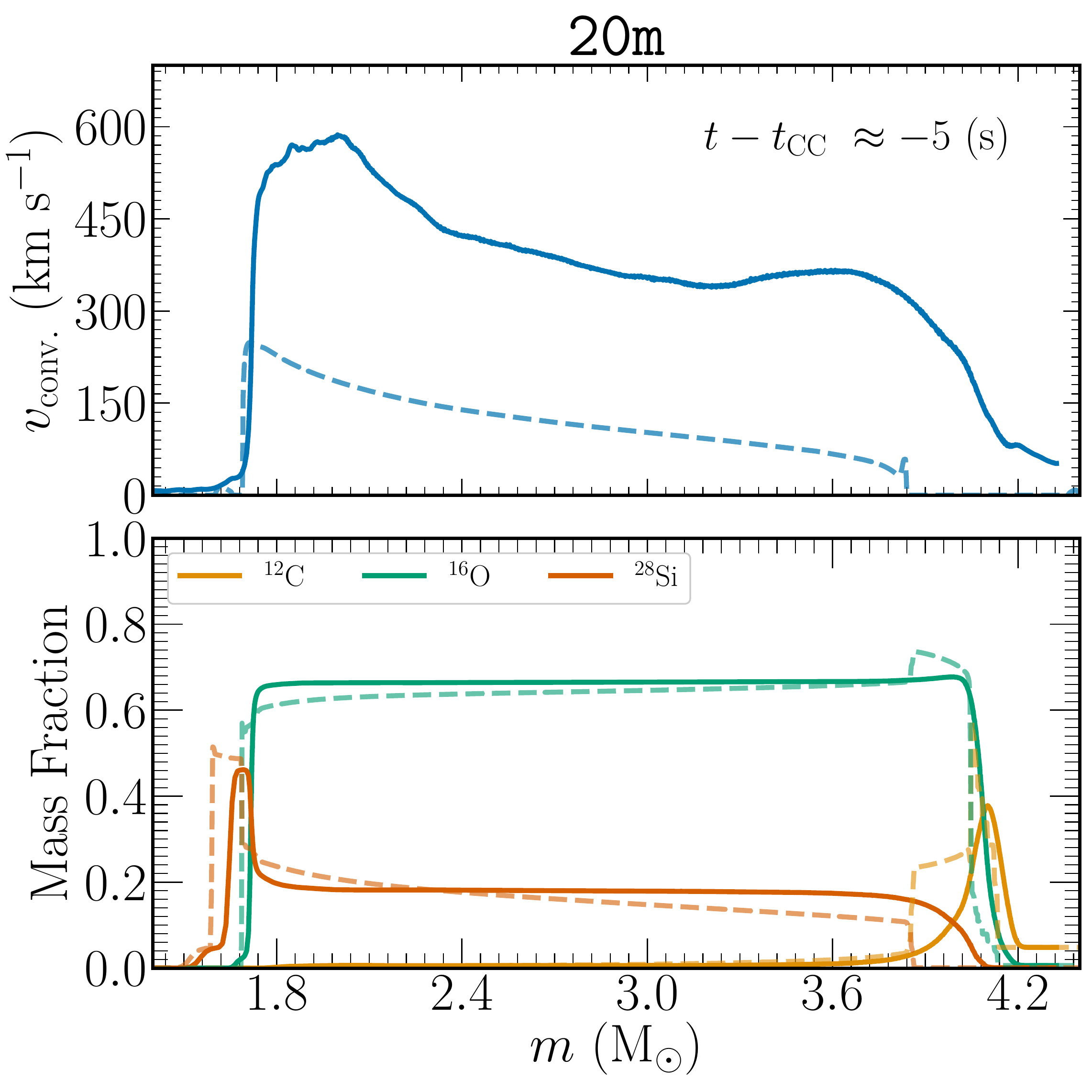}}
        \end{subfigure}
        \caption{Same as in Figure~\ref{fig:14m_compare_mesa} but for the corresponding 20 \msun models.
        }\label{fig:20m_compare_mesa}
\end{figure}

\subsubsection{14 \msun}
In Figure~\ref{fig:14m_compare_mesa} we show the angle-average tangential velocity profile (top) and 
the specific mass fraction for three major isotopes: $^{12}$C, $^{16}$O, and $^{28}$Si (bottom) for the 3D \texttt{14m}
model approximately five seconds prior to core-collapse. Also shown are the corresponding \MESA profiles
at the same time before core-collapse denoted by the dashed lines of the same color. 
For the 14 \msun model, we find that a few qualitative features worth mentioning. 
First, the 1D \MESA model shows a peak convective velocity in the O-shell region of $v_{\rm{conv.}}\approx$ 50
km s$^{-1}$. This value is approximately \emph{four times} smaller than predicted by our 3D tangential velocity
profiles where a peak value of velocity is found of $v_{\rm{conv.}}\approx$ 210 km s$^{-1}$. Beyond this, we observe that the
1D model is more compact with the shell locations closer to the iron core (lower specific mass coordinate). This difference
is likely attributed to the slight expansion of the 3D model (See Figure~\ref{fig:heatmaps_v_conv}, top right). 

The mass fraction profiles between the 1D and 3D models are qualitatively similar other than the 3D model being less compact 
(shells at larger mass coordinate) and the compositional gradients being smoothed at the boundary due to increased 
mixing. One notable feature in the 3D model is the lack of a Si-rich region at the base of the Si-shell 
region. The 1D model shows a peak in $^{28}$Si from $m\approx$ 1.50-1.54 \msun. However, in the 3D simulation, 
we observe instead one merged smoothed Si region. This merged region is able to reach higher tangential velocity 
speeds of $v{\rm{conv.}}\approx$ 80 km s$^{-1}$ in the 3D model. 

\subsubsection{20 \msun}
Figure~\ref{fig:20m_compare_mesa} shows the convective velocity and mass fraction profiles for the 
20 \msun 3D and 1D \MESA models at the same time as in Figure~\ref{fig:14m_compare_mesa}. We 
find that the location of the respective shell locations between 1D and 3D are in better agreement than the 
14 \msun models which showed a difference of $\Delta M_{\rm{O}} \approx$ 0.02 \msun.
In the 20 \msun models compared here, the location of the base of the 
O-shell is found at a specific mass coordinate of $m\approx$ 1.8\msun.
The extent of the O-shell differs slightly between the two model, likely attributed to the mixing at the boundary
between the O- and C-shell regions. The extra mixing at this boundary will be discussed further in \S~\ref{sec:results_20m}.
Similar to the 14\msun model, the convective velocity in the 1D model is less than observed in the 3D simulation. In this 
particular case, we find speeds three times larger in the 3D model than what is predicted by the 1D \MESA model.  
The mass fractions for both models follow similar behavior as the 14 \msun models (but with less shell expansion / contraction)
in that the profiles are smoothed out and sharp features from the 1D model are not present. One particular case is 
at the edge of the O-shell region where a slight increase of $^{12}$C and $^{16}$O from $m\approx$ 3.9-4.2 \msun
is not observed in the 3D simulation, suggesting it is mixed in or out of the O-shell region during the simulation in 3D.

\begin{figure}[!t]
         \centering  
        \begin{subfigure}{
                \includegraphics[width=0.47\textwidth]{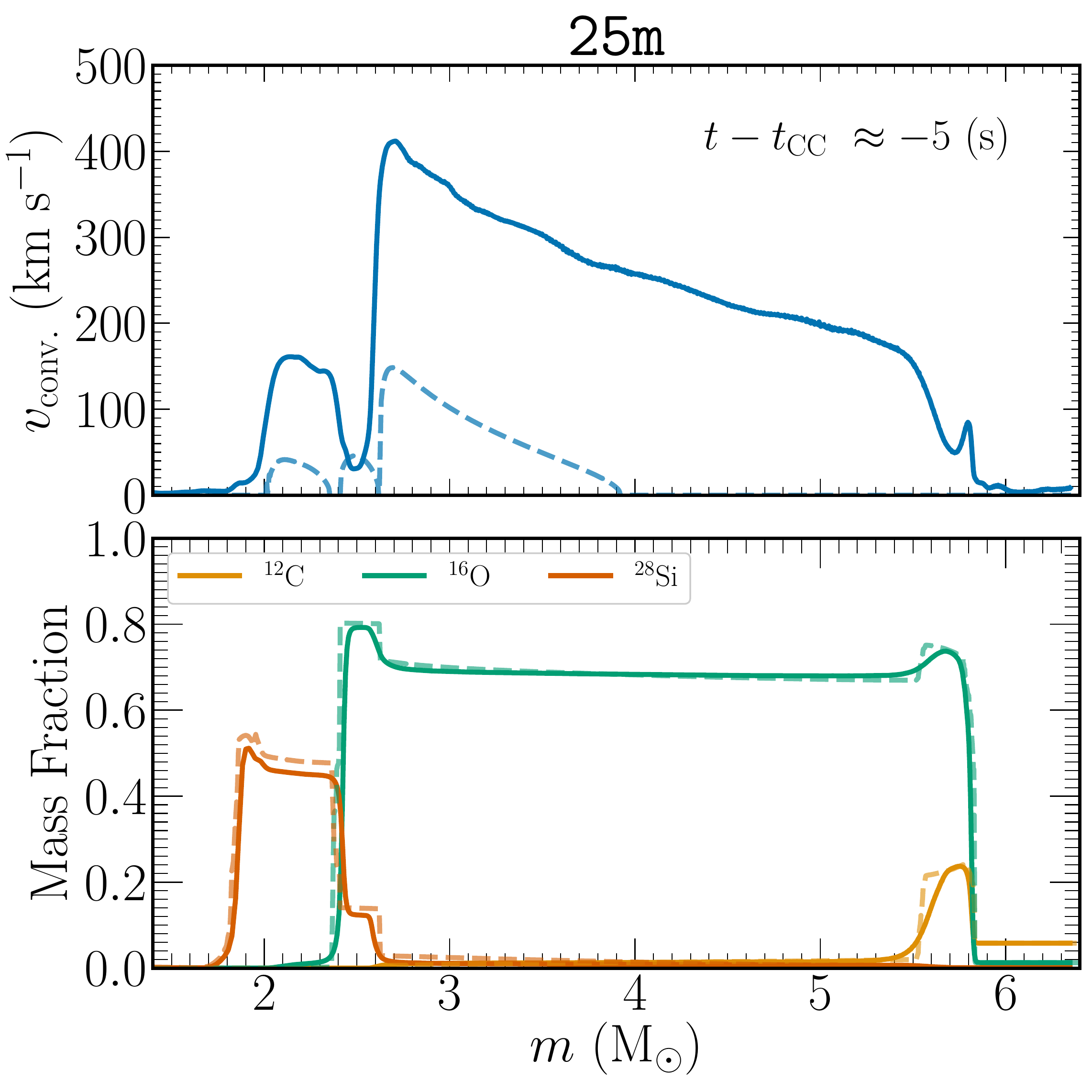}}
        \end{subfigure}
        \caption{Same as in Figure~\ref{fig:14m_compare_mesa} but for the corresponding 25 \msun models.
        }\label{fig:25m_compare_mesa}
\end{figure}

\subsubsection{25 \msun}

Lastly, the convective and isotopic mass fractions profiles for the 25-\msun models are shown in Figure~\ref{fig:25m_compare_mesa} 
at $t-t_{\rm{CC}}=-5$ s. In the 25-\msun model, we see similar shell location agreement between 
the 3D and 1D \MESA models suggesting less expansion/contraction compared to the 14\msun model and the 
15\msun model of FC20. However, unlike the 20 \msun model, and to some degree the 14\msun model, 
the extent of the O-shell region disagrees significantly between the 3D and 1D models. The 3D 25 \msun model
suggests a convective O-shell region that extends from $m\approx$ 2.5-6.0 \msun where and the \MESA model
has the convective speeds going to zero at $m\approx$ 4.0 \msun. In the bottom panel, the $^{16}$O mass fraction
profiles agree remarkably aside from the smoothed composition gradients in the 3D profile. We find tangential velocity
speeds in the O-shell region that peak at $v_{\rm{conv.}}\approx$ 400 km s$^{-1}$, a factor of four times larger than 
predicted by the \MESA model at the base of the O-shell. Recall that in the Si-shell region of the 1D \MESA model at this
time we see two convective shells with speeds of $v_{\rm{conv.}}\approx$ 30 km s$^{-1}$. In the 3D model, we observe 
instead a merged Si-shell with $v_{\rm{tan.}}\approx$ 260 km s$^{-1}$.

\begin{figure}[!t]
\centering{\includegraphics[width=1.0\columnwidth]{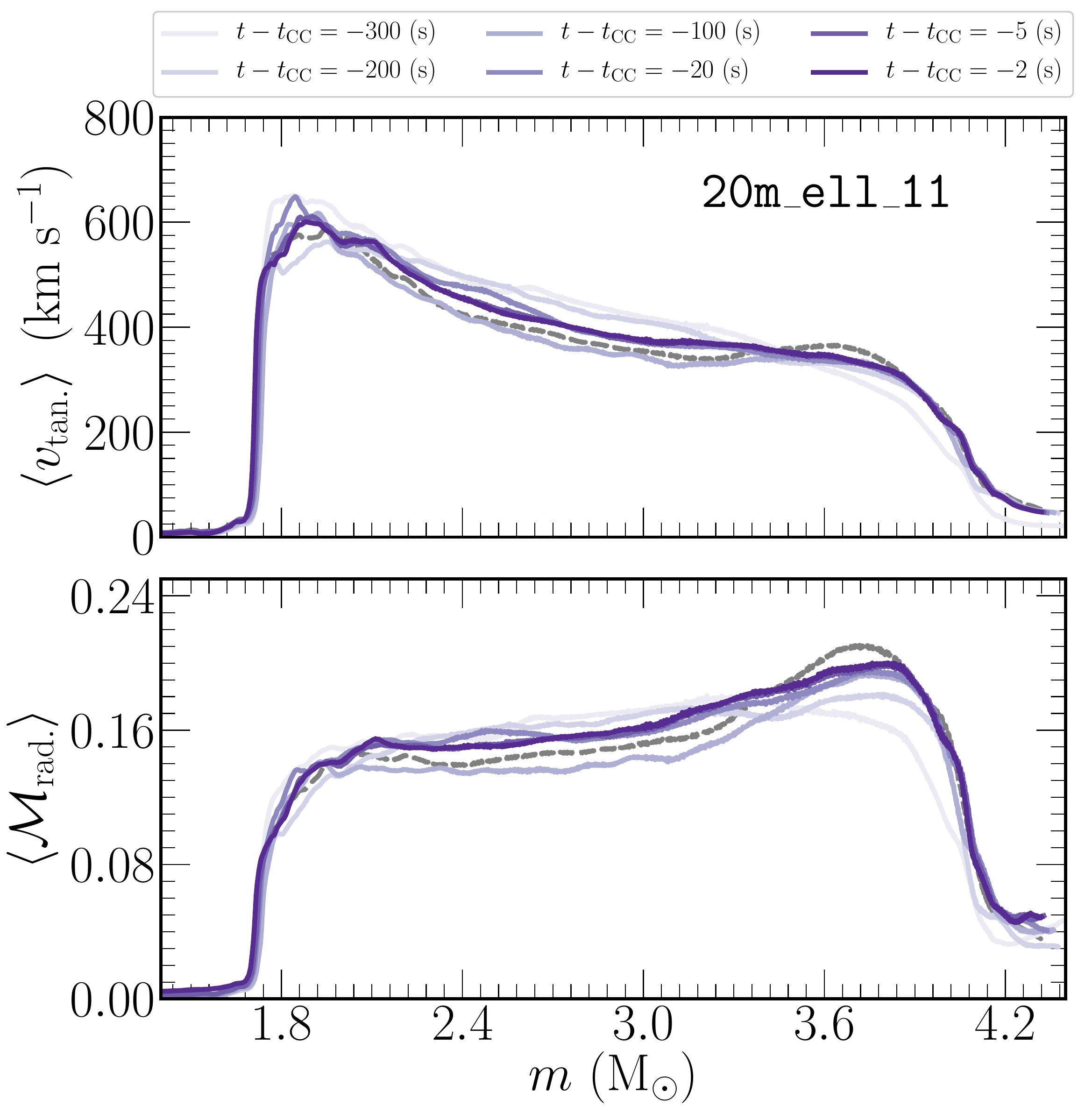}}
\caption{Same as in Figure~\ref{fig:3d_14m_conv_props} but for the $\texttt{20m\_ell\_11}$ model. 
The dashed gray line are the profiles from the $\texttt{20m}$ at $t-t_{\rm{CC}}=-2$ s plotted again here for comparison.
}\label{fig:3d_20m_ell_11_conv_props}
\end{figure}

\begin{figure}[!htb]
\centering{\includegraphics[width=1.0\columnwidth]{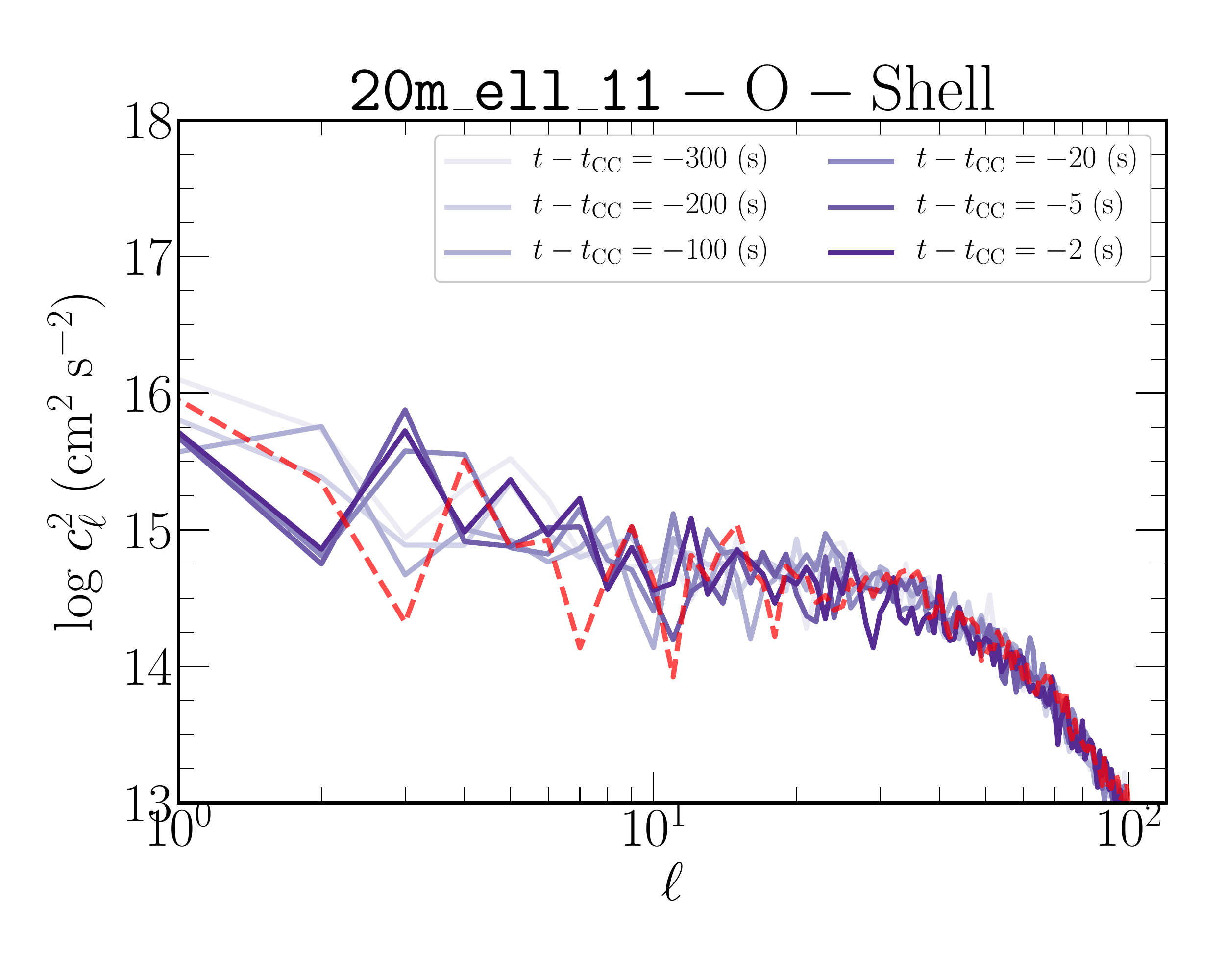}}
\caption{Same as in Figure~\ref{fig:3d_power_spectra} but for the \texttt{20m\_ell\_11} 3D model. Also shown 
is the spectrum for the \texttt{20m} model at $t-t_{\rm{CC}}=-5$ s (red dashed line) for comparison.
}\label{fig:3d_power_spectra_20m_o_ell_11}
\end{figure}

\begin{figure*}[!htb]
\centering{\includegraphics[width=2.0\columnwidth]{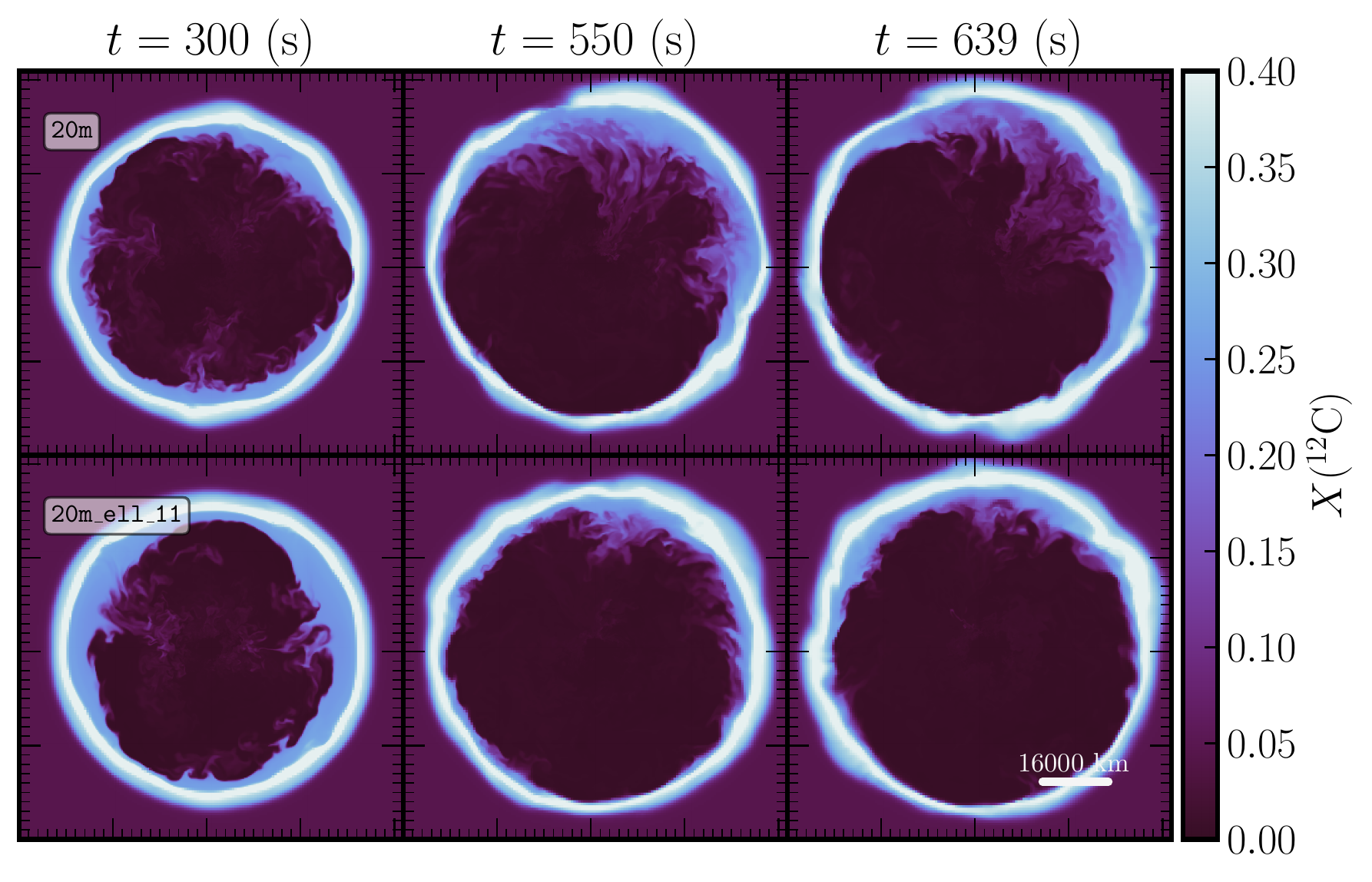}}
\caption{Slice in the $x-z$ plane of the $^{12}$C mass fraction for the two 3D 20 \msun models at $t=300$ s, 
$t=550$ s, and $t=639$ s (5 seconds before collapse) from left to right, respectively. 
}\label{fig:3d_c12_slice_y}
\end{figure*}

\subsection{A closer look at the 20 \msun models}
\label{sec:results_20m}
In order to test the effect of our choice of velocity field initialization methods described in \S~\ref{sec:methods},
we explore the results of our additional 3D 20 \msun model. This model was evolved using a larger spherical harmonic 
index for the O-shell ($\ell=11$) than the fiducial 20-\msun model resulting in initially smaller scale perturbations. Otherwise, 
the quantities used to initialize the velocity field are the same as listed in Table~\ref{tbl:initial_models} for \texttt{20m}. We 
label this additional 3D 20 \msun model \texttt{20m\_ell\_11}. Figure~\ref{fig:3d_20m_ell_11_conv_props} shows the
angle-average tangential velocity and Mach number for this model at six times prior to collapse. Also shown are
are the profiles from the $\texttt{20m}$ plotted again here for comparison via thin dashed lines. The differences
between the profiles of the two models are quantitively small. The main differences can be observed near the 
edge of the O-shell region, $m\approx$ 3.8 \msun where the \texttt{20m} model (dashed line) shows a \emph{slightly}
larger velocity and radial Mach number than the \texttt{20m\_ell\_11} model. We can more quantitatively explore
the differences between these models by considering the power spectrum of the O-shell region.

We compute the power spectrum for the O-shell region for the \texttt{20m\_ell\_11} in
Figure~\ref{fig:3d_power_spectra_20m_o_ell_11}.  For this model, we compute the spectra in 
the same way as the $\texttt{20m}$, taking $r_{\rm{O}}=5000$ km. The spectrum five seconds prior to collapse 
for the \texttt{20m} is also shown for comparison (red dashed line). The spectra shown are qualitatively similar 
except for the \texttt{20m\_ell\_11} model showing the dominant mode to be at $\ell=3$ near collapse 
(instead of $\ell=1$ for the \texttt{20m} model). At $t-t_{\rm{CC}}=-5$ c, the \texttt{20m} model shows slightly 
more power at $\ell=1$ and $\ell=2$ with a deficit of power at $\ell=3$ compared to the other model. At 
intermediate scales, the models show relatively similar power.

\begin{figure}[!t]
\centering{\includegraphics[width=1.0\columnwidth]{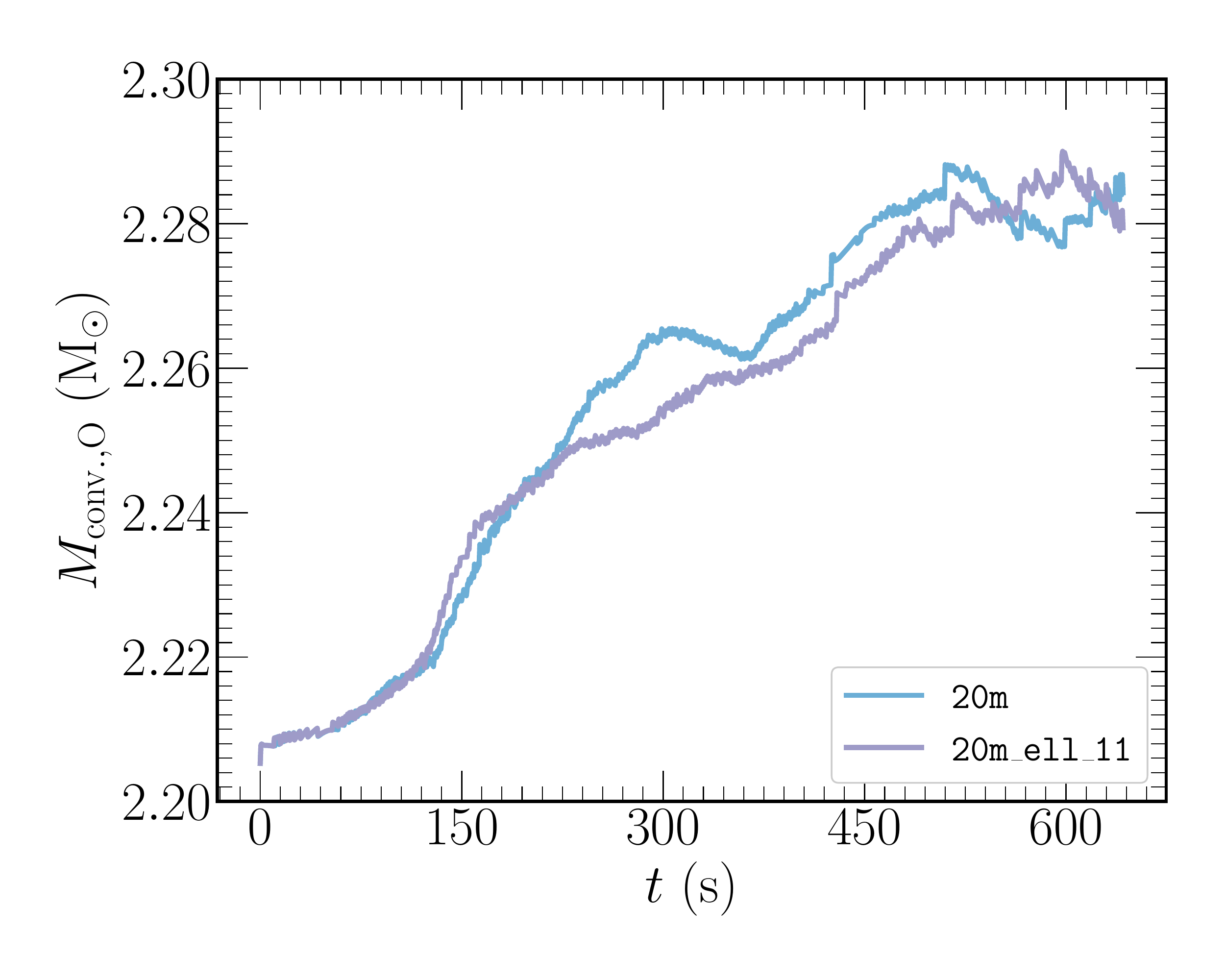}}
\caption{Mass of the convective O-shell region as a function of time for the \texttt{20m} and \texttt{20m\_ell\_11} models.
}\label{fig:3d_20m_mass_entrainment_rate}
\end{figure}

In Figure~\ref{fig:20m_compare_mesa} we observed smoothing of the $^{12}$C chemical stratification in the 
3D angle-average profiles that were not observed in \MESA. This smoothing suggests additional mixing 
at the boundary of the convective O-shell region. Instabilities at the convective boundaries can induce mixing 
described as turbulent entrainment. In the case of O-shell burning, such entrainment can lead to ingestion of 
C fuel into the O-shell, the likes of which could have implications for the synthesis of odd-$Z$ elements depending 
on the entrainment rate \citep{ritter_2018_aa}. Turbulent entrainment has been studies previously in 3D 
simulations of convection O-shell burning \citep{meakin_2007_ab, muller_2016_aa,andrassy_2020_aa}. 
Here, we explore the mixing at the boundary of the C/O shell regions and the dependence of these results on our 
choice of initial perturbations in our two 20 \msun models.

We can define the bulk Richardson number as a measure of the relative stiffness of the C/O shell boundary.
Using a form of the equation that relates the convective scale length to that of the pressure scale height we can 
define the bulk Richardson number as 
\begin{equation}
R_{\rm{B}} = \frac{\delta \rho}{\rho} \frac{P}{\rho v^{2}_{\rm{conv.}}}~,
\end{equation}
where $\delta \rho / \rho$ corresponds to the density contrast across the boundary and $v_{\rm{conv.}}$ refers 
to the turbulent velocity components perpendicular to the boundary. For the \texttt{20m} model we use $\delta \rho / \rho \approx0.22$ 
and an convective velocity of $v_{\rm{conv.}}\approx$ 320 km s$^{-1}$ near the boundary, both computed using 
angle-average profiles of the relevant quantity. Using these values 
we find a dimensionless bulk Richardson number of $R_{\rm{B}}\approx$ 37. This result suggests a relatively ``soft'' 
boundary that is slightly more stiff than the 18 \msun model seen in \citet{muller_2016_aa} and similar to the core convection 
boundary in \citet{meakin_2007_ab}. The \texttt{20m\_ell\_11} model only shows a slight difference with $R_{\rm{B}}\approx$ 41.

\begin{figure}[!t]
         \centering  
       \begin{subfigure}{
                \includegraphics[width=0.48\textwidth]{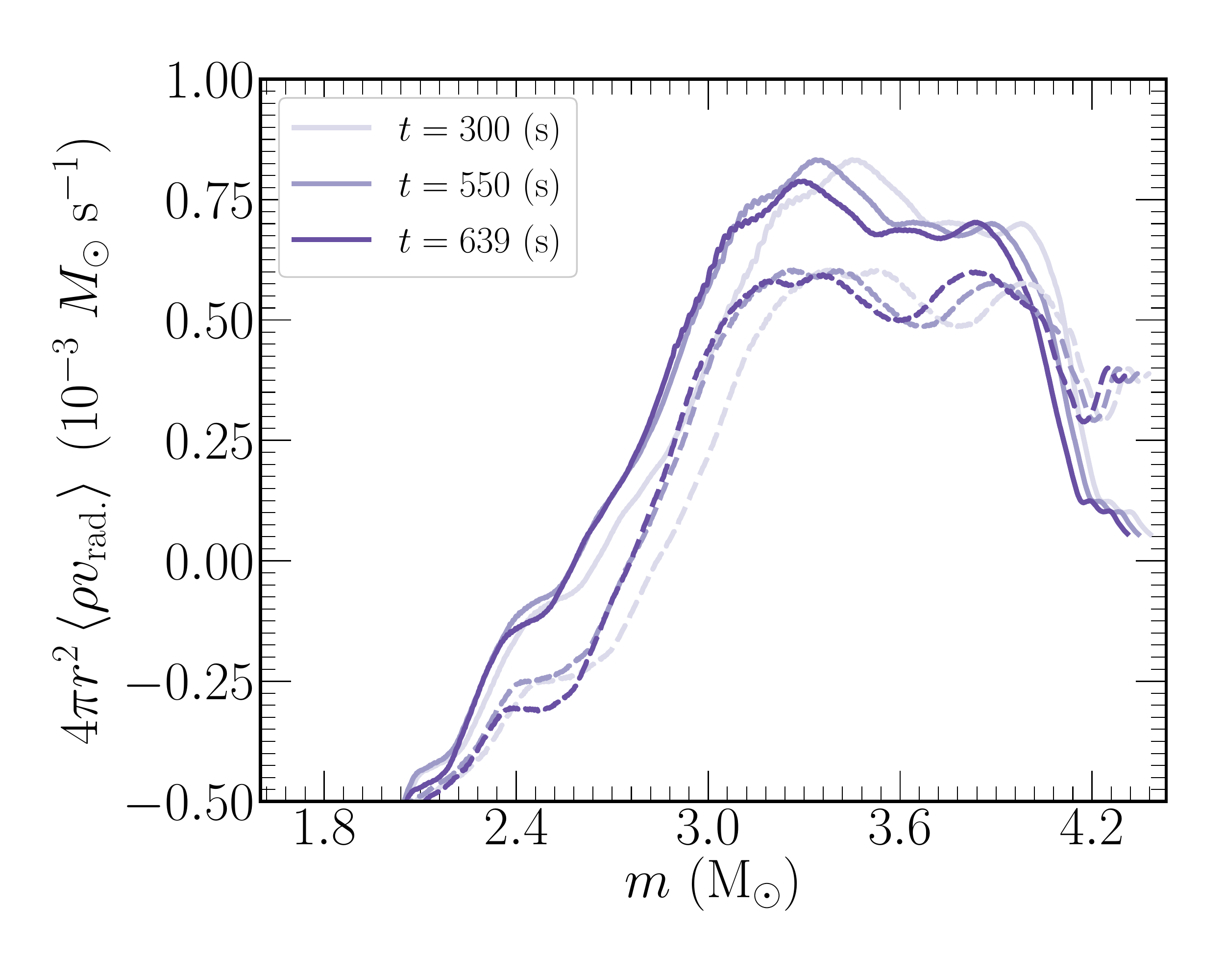}}
        \end{subfigure}
        \caption{Turbulent mass flux as a function of specific mass coordinate for the two 3D 20 \msun models (model \texttt{20m\_ell\_11} is shown by the dashed lines) at $t=$ 300, 550, and 639 s.
        The approximate location of the C/O interface corresponds to a specific mass coordinate of $m\approx$ 4 \msun.
        }\label{fig:3d_20m_mass_flux}
\end{figure}

We show slice plots of the $^{12}$C mass fraction for both 3D models at three different times near collapse in Figure~\ref{fig:3d_c12_slice_y}. Both models show relatively similar distributions of $^{12}$C at $t=300$ s primarily 
influenced by the interaction of the initial velocity field produced by our perturbations and the shell boundary. The 
subsequent aspherical deformation of this boundary begins to diverge between the two models due to the different drive 
scale $\ell$ that characterizes the velocity field. Qualitatively,  \texttt{20m} shows a larger fraction of $^{12}$C
being pulled down in the northern hemisphere of the slice of model. The bottom panel suggest that less $^{12}$C
is being entrained and that this is potentially linked to the choice of $\ell$ in the initialization of the velocity field. 

\begin{figure*}[!htb]
         \centering  
        \begin{subfigure}{
                \includegraphics[width=\textwidth]{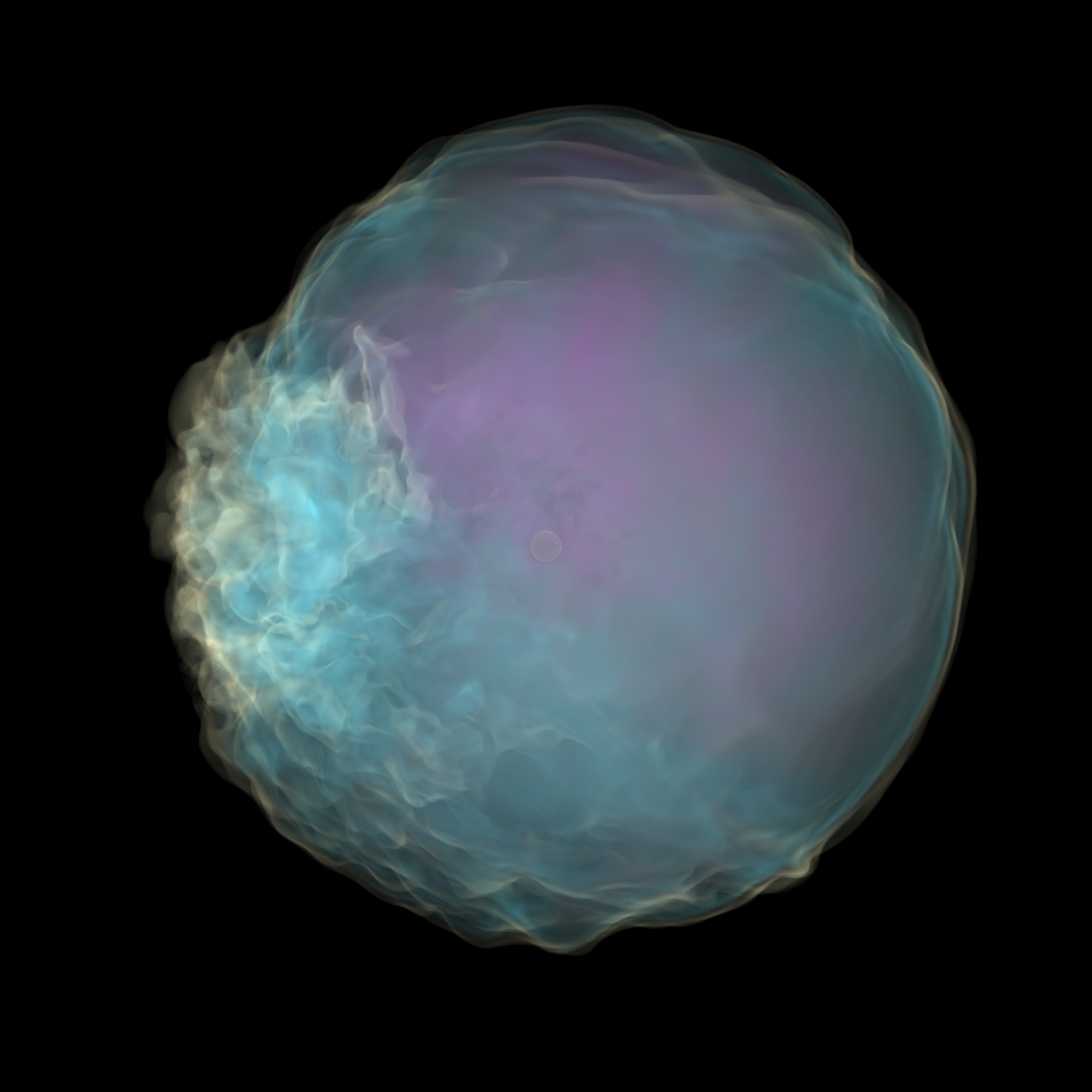}     
                }
        \end{subfigure}
        \caption{Volume rendering of the $^{28}$Si mass fraction for the \texttt{20m} model at $t=$ 639 s. The transfer function consists of 
        three Gaussian peaks to highlight the low (yellow, 0.05), moderate (blue, 0.1), and high (purple, 0.2) concentration of Si. The 
        low concentration regions highlight fresh $^{12}$C fuel being mixed down into the O-shell region. The blue region trace
        the approximate extent and structure of the convective O-shell region. Purple regions follow high concentrations of fresh Si
        being mixed due to O-shell burning. The iron core at this point spans an approximate radius of $r\approx$ 2520 km and is shown in gold at the center of the model. The O-shell at this time spans a radius of $r\approx$ 60 000 km. This visualization was produced 
        using \texttt{yt} and the colormap library \texttt{cmocean}.
        }\label{fig:3d_20m_vol_si28}
\end{figure*}

To explore this more quantitatively, we compute the mass of the convective O-shell as a function of time for both models.
In Figure~\ref{fig:3d_20m_mass_entrainment_rate} we show the approximate mass of the convective O-shell 
region for both of our 20 \msun 3D models. We define this to be the mass enclosed between an angle-average specific 
entropy of $s=[3.6,6.1]$ $k_{\rm{B}}$ baryon$^{-1}$ When considering the temporal evolution of the mass of the 
convective O-shell, we can observe the same divergence in the mass at times greater than $t\approx$ 240 s. Beyond 
this point, the mass increases at a faster rate than the other 20 \msun model before reaching a saturation point at 
$t\approx$ 375 s, where both models follow a similar slope. In the last $\approx$ 100 s, the \texttt{20m} model 
shows a slight decrease in mass due to the deformation of the edge of the O-shell leading to fluctuations in the 
average shell radius. 

In Figure~\ref{fig:3d_20m_mass_flux} we show mass flux in the O-shell region of our 20 \msun models (model \texttt{20m\_ell\_11} is shown by the dashed lines) at 
$t=$ 300, 550, and 639 s. In the \texttt{20m} model we observe a positive flux at the boundary of 
the C/O shell region ($m\approx$ 4 \msun)
of 7.5 $\times$ 10$^{-4} $ \msun s$^{-1}$. The \texttt{20m\_ell\_11} model shows a \emph{slightly} less efficient
mass entrainment 5.2 $\times$ 10$^{-4} $ \msun s$^{-1}$. Despite having a more stiff boundary between the C/O
shell regions in our simulations, we observe a larger rate of mass entrainment than found in \citet{muller_2016_aa}. 
Our results also 
suggest that the characteristic properties of the velocity field initialization can lead to slight differences in the 
net mass entrainment rate as observed in our two 20 \msun simulations. Larger scale initial perturbations 
can lead to more significant deformation of the overlying C-shell region which can enhance (potentially artificially) the
ingested $^{12}$C into the O-shell region. In Figure~\ref{fig:3d_20m_vol_si28} we show a volume rendering of the
$^{28}$Si mass fraction for the \texttt{20m} model at $t=639$ s. The yellow and blue regions trace the low and 
moderate concentration of $^{28}$Si. These regions trace mixing of the $^{12}$C into the O-shell region. The iron
core is shown by a gold spherical contour with a sharp peak Gaussian at an approximate entropy $s=$ 4 $k_{\rm{B}}$ baryon$^{-1}$.

\section{Summary and Discussion}
\label{sec:discussion}
We have presented 3D hydrodynamic simulations of O- and Si-shell burning in massive star models 
using \texttt{MESA} and the \texttt{FLASH} simulation framework. We follow up to the final ten minutes prior to 
core-collapse to capture the development of the turbulent convective flow prior to and including gravitational
collapse. In this study, we considered initial 1D progenitor models of 14-, 20-, and 25 \msun 
to survey a range of O/Si shell density and compositional configurations. We also evolved an additional 
3D 20 \msun model to investigate the impact of our choice of initial velocity perturbations. 

In our 14 \msun model, we observed relatively weak O-shell convection with the peak of the 
power spectrum near a spherical harmonic index of $\ell=4$. 
We also observed 
a slight increase in 
power at larger scales $\ell=1-3$ near collapse for this model. The convective O-shell region 
showed radial Mach numbers of $\mathcal{M}_{\rm{rad.}}\approx0.05$ near collapse.
Despite weaker convection 
compared to our other 3D models, we found that angle-average convective velocity profile was 
approximately four times larger than the speeds predicted by \texttt{MESA} in the moments 
prior to collapse for our choice of mixing length parameters.

Our baseline \texttt{20m} model showed the most energetic power spectrum in the O-shell region 
with power residing at the largest scales of $\ell=1-3$ near collapse. The convective velocity profile 
showed speeds \emph{three} times larger than the 1D \texttt{MESA} model counterpart. 
We also simulated an additional 20 \msun model with a different initial velocity field topology in the O-shell to 
explore the impact of our initiial conditions. When choosing an initially smaller scale ($\ell=11$ instead of $\ell=7$) 
velocity field topology we observed variations in the amount of turbulent entrainment of $^{12}$C at the 
C/O shell interface. The model with smaller scale topology for the O-shell velocity field (model \texttt{20\_ell\_11}) 
showed a reduced positive turbulent mass flux at the C/O interface and an overall less efficient mass 
entrainment rate of 5.2 $\times$ 10$^{-4} $ \msun s$^{-1}$ near collapse. This value was found to be 
about $\approx$ 31 \% less than the baseline 20 \msun model (using $\ell=7$) which showed prominent 
large-scale mixing at the boundary leading to additional C-ingestion at a time $t-t_{\rm{CC}}=-300$ s.
Both of our 3D 20 \msun models found Mach numbers of $\mathcal{M}_{\rm{rad.}}\approx0.15-0.20$ 
in the convective O-shell region maintained throughout the final 300 seconds prior to collapse.

The results of our 25 \msun showed 
qualitatively properties similar to the 14- and 20 \msun model. At early
times, the power spectrum shows a peak near $\ell=4$. However, at later times, energy 
is transferred towards larger scales with the bulk of energy at $\ell=1-3$ near collapse. In this model, the 
convectively active Si-shell region can be characterized by power over a broad range of intermediate
scales of $\ell=10-20$ as the simulation approaches collapse. This model showed radial Mach numbers of $\mathcal{M}_{\rm{rad.}}\approx0.03$ and $\mathcal{M}_{\rm{rad.}}\approx0.08$ in the Si- and O-shell regions in the 
seconds prior to collapse, respectively.

The results of the 3D simulations presented 
in this work show a discrepancy between angle-average profiles of the convective velocities and the 
1D \MESA convection profiles for our choice of input mixing length parameters. This discrepancy has 
been seen in previous studies.
In \citet{jones_2017_aa} they performed simulations of idealized $4\pi$ 3D O-shell burning
of a 25\msun model. The results of this study showed that 
the convective velocities in the MLT framework, when using diffusion coefficients computed from the 
angle-average 3D data, can agree within a factor of a few compared to predictions by MLT and the 1D \MESA model. 
Work by \citet{herwig_2006_aa} found convective speeds predicted by MLT were a few factors slower 
than their 2D hydrodynamic simulations of He-flash convection. 
The discrepancies presented here can be attributed in part to the fact that the convective and burning timescales become similar 
in the moments leading towards collapse. MLT assumes that the convective mixing happens on the local dynamical timescale and 
is much shorter than other relevant timescales. Moreover, MLT is assumes a mixing length over which an average 
velocity is computed, as opposed to a range of convective \emph{time-dependent} scales in the radial and
transverse directions as presented in the 3D models of this work. Efforts to provide a 
new formulation that takes these assumptions and other aspects of turbulent convection into account for a new 1D prescription of stellar convection has been considered in \citet{arnett_2015_aa}. Improvements to MLT 
have also been proposed \citep{canuto_1991_aa,trampedach_2014_aa}.
 
Recent work by \citet{yoshida_2021_aa} explore the final minutes prior to collapse in 3D simulations of 
Si- and O-shell burning in 22\msun and 27\msun progenitors. In their models, they observe peak 
spherical harmonic modes of $\ell_{\rm{peak}}=2-3$ and find that the Mach number
can exceed 0.1. These results agree in general with the values found for our 20\msun and 25\msun models. 
Except for the case of the radial Mach number of the 25\msun model where we find that the 
peak value near collapse is only found to be $\approx$ 0.08. They also observe episodic burning burning 
of O and Ne which lead to an increase in the turbulent mixing found in their simulations. Similar 
episodic burning was found in the 18.88 \msun model of \citet{yadav_2019_aa}. We find no evidence 
of episodic burning in any of our four 3D simulations.

The set of models presented in this work are a step forward in efforts to produce realistic 3D pre-supernova 
models that capture the properties of massive stars in their final moments prior to collapse. A key 
component not yet discussed in this paper is the effect of rotation and magnetism. 
Recent work by \citet{muller_2020_aa} suggests that pre-SN models with slow to 
moderately rotating cores near collapse could play a role in the delayed neutrino-driven mechanism of 
CCSNe. Efforts towards addressing the impact of magnetic fields in 3D simulations of O-shell burning 
were performed recently for a 18 \msun model, although this model did not include the effect of 
rotation \citep{varma_2021_aa}. Rotation and magnetism must be considered in 
realistic 3D progenitor models as their inclusion is relevant to magnetically-driven \citep{mosta_2015_aa,fryer_2019_aa} and
ordinary neutrino-driven CCSN explosions \citep{curtis_2020_aa,barker_2021_aa} and the multi-messenger signals they produce \citep{warren_2020_aa,pajkos_2021_aa}.

\software{
\MESA \citep[][\url{http://mesa.sourceforge.net}]{paxton_2011_aa,paxton_2013_aa,paxton_2015_aa,paxton_2018_aa},
\FLASH \citep[][\url{http://flash.uchicago.edu/site/}]{fryxell_2000_aa},
\texttt{yt} \citep[][\url{https://yt-project.org}]{turk_2011_aa}, and
\texttt{matplotlib} \citep[][\url{https://matplotlib.org}]{hunter_2007_aa}.}

\acknowledgements
We thank 
Josh Dolence,
Sam Jones,
Brian O'Shea,
and
P. N. Sagan,
for useful discussions. 
C.E.F. acknowledges support from the National Science Foundation Graduate Research Fellowship 
Program under grant number DGE1424871. Research presented in this article was supported by the 
Laboratory Directed Research and Development program of Los Alamos National Laboratory under 
project number 20210808PRD1. 
S.M.C. is supported by the U.S. Department of Energy, Office of Science, Office 
of Nuclear Physics, Early Career Research Program under Award Number DE-SC0015904. 
This material is based upon work supported by the U.S. Department of Energy, Office of 
Science, Office of Advanced Scientific Computing Research and Office of Nuclear Physics, 
Scientific Discovery through Advanced Computing (SciDAC) program under Award Number 
DE- SC0017955. This research was supported by the Exascale Computing Project (17-SC-20-SC), 
a collaborative effort of the U.S. Department of Energy Office of Science and the National Nuclear 
Security Administration.
This work was supported in part by Michigan State University through 
computational resources provided by the Institute for Cyber-Enabled Research.
This research made extensive use of the SAO/NASA Astrophysics Data System (ADS).

\appendix

\section{\texttt{FLASH} Simulation Resolution}
\label{sec:app1}

\begin{figure}[!htb]
         \centering  
        \begin{subfigure}{
                \includegraphics[width=0.5\textwidth]{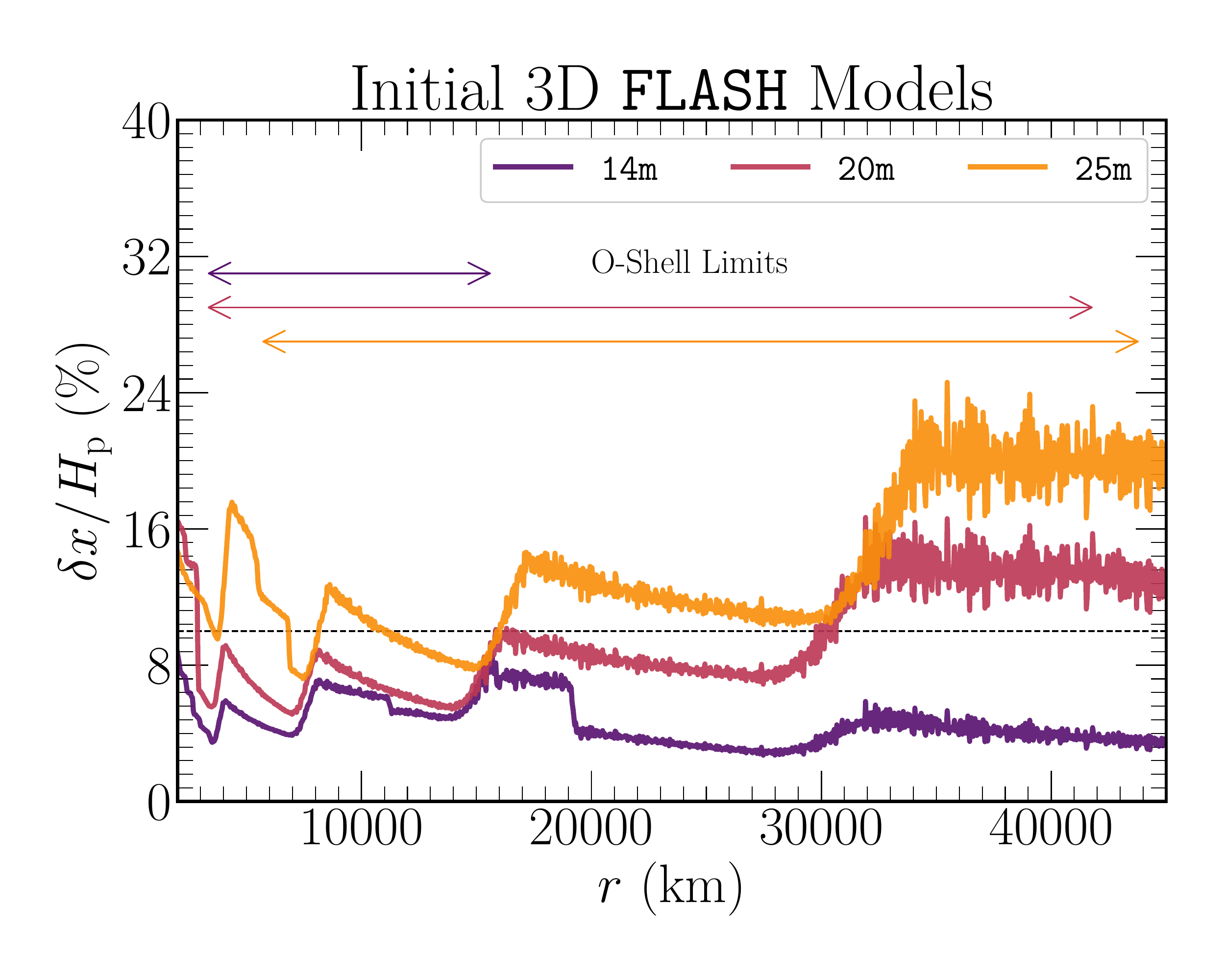}}
        \end{subfigure}
        \caption{The angle-average cell resolution as a percentage of the relative pressure scale height 
        for the three 3D models at $t=0$. Also shown are the approximate O-shell radial limits. The dashed
        horizontal line corresponds to 10\% of the pressure scale height.
        }\label{fig:compare_all_models_res}
\end{figure}

The finest level of refinement for each model results in a grid spacing of $\approx$ 32.5 km. 
The approximate effective resolution for each model varies due to their different initial shell configurations. 
For the 14 $M_{\odot}$ model, the finest resolution level is situated at the center of the simulation and 
extends out to a radius of $r\approx$ 3500 km. This region includes the entire iron core and Si-shell resulting
in an average resolution of $\approx$ 8 $\%$ of the local pressure scale height, $H_{\rm{p}}= P / \rho G$. The 
resolution decreases beyond this radius by a factor of two out to $r\approx$ 7100 km. In this region, our simulation
has a resolution that equates to $\approx$ 7$\%$ $H_{\rm{p}}$. The resolution continues to decrease beyond
this radius based based on logarithmic changes in specific density, pressure, and velocity. After initialization, 
the location of the refinement levels do not change throughout the simulation. The 20 $M_{\odot}$ model has 
a qualitatively similar grid with the finest resolution level containing the entire iron and Si-shell region. In the 25 $M_{\odot}$, 
model, the finest refinement level extends out to $r\approx$ 4100 km. Unlike the other models, in the 25 $M_{\odot}$ model,  
this only encompasses
the iron core and a portion of the Si-shell region. Within this finest refinement level, the resulting resolution 
equates to an average of $\approx$ 5$\%$$H_{\rm{p}}$. Beyond this radius, the grid resolution is 
decreased by a factor of two but the simulation maintains an approximate effective resolution of 
$\approx$ 5$\%$$H_{\rm{p}}$ out to $r\approx$ 8100 km. The remainder of the grid resolution is similar to the 
other two models. In Figure~\ref{fig:compare_all_models_res} we show the average cell reolution as a percentage 
of the pressure scale height according to the 3D models. Also shown are the approximate O-shell radial 
limits for each model.

\bibliographystyle{aasjournal}
\bibliography{prog3dn}

\begin{thebibliography}{}
\expandafter\ifx\csname natexlab\endcsname\relax\def\natexlab#1{#1}\fi
\providecommand{\url}[1]{\href{#1}{#1}}
\providecommand{\dodoi}[1]{doi:~\href{http://doi.org/#1}{\nolinkurl{#1}}}
\providecommand{\doeprint}[1]{\href{http://ascl.net/#1}{\nolinkurl{http://ascl.net/#1}}}
\providecommand{\doarXiv}[1]{\href{https://arxiv.org/abs/#1}{\nolinkurl{https://arxiv.org/abs/#1}}}

\bibitem[{{Andrassy} {et~al.}(2020){Andrassy}, {Herwig}, {Woodward}, \&
  {Ritter}}]{andrassy_2020_aa}
{Andrassy}, R., {Herwig}, F., {Woodward}, P., \& {Ritter}, C. 2020, \mnras,
  491, 972, \dodoi{10.1093/mnras/stz2952}

\bibitem[{{Arnett} {et~al.}(2015){Arnett}, {Meakin}, {Viallet}, {Campbell},
  {Lattanzio}, \& {Moc{\'a}k}}]{arnett_2015_aa}
{Arnett}, W.~D., {Meakin}, C., {Viallet}, M., {et~al.} 2015, \apj, 809, 30,
  \dodoi{10.1088/0004-637X/809/1/30}

\bibitem[{{Barker} {et~al.}(2021){Barker}, {Harris}, {Warren}, {O'Connor}, \&
  {Couch}}]{barker_2021_aa}
{Barker}, B.~L., {Harris}, C.~E., {Warren}, M.~L., {O'Connor}, E.~P., \&
  {Couch}, S.~M. 2021, arXiv e-prints, arXiv:2102.01118.
\newblock \doarXiv{2102.01118}

\bibitem[{{Burrows} {et~al.}(2020){Burrows}, {Radice}, {Vartanyan}, {Nagakura},
  {Skinner}, \& {Dolence}}]{burrows_2020_aa}
{Burrows}, A., {Radice}, D., {Vartanyan}, D., {et~al.} 2020, \mnras, 491, 2715,
  \dodoi{10.1093/mnras/stz3223}

\bibitem[{{Canuto} \& {Mazzitelli}(1991)}]{canuto_1991_aa}
{Canuto}, V.~M., \& {Mazzitelli}, I. 1991, \apj, 370, 295,
  \dodoi{10.1086/169815}

\bibitem[{{Couch} {et~al.}(2015){Couch}, {Chatzopoulos}, {Arnett}, \&
  {Timmes}}]{couch_2015_aa}
{Couch}, S.~M., {Chatzopoulos}, E., {Arnett}, W.~D., \& {Timmes}, F.~X. 2015,
  \apjl, 808, L21, \dodoi{10.1088/2041-8205/808/1/L21}

\bibitem[{{Couch} {et~al.}(2013){Couch}, {Graziani}, \&
  {Flocke}}]{couch_2013_ab}
{Couch}, S.~M., {Graziani}, C., \& {Flocke}, N. 2013, \apj, 778, 181,
  \dodoi{10.1088/0004-637X/778/2/181}

\bibitem[{{Couch} \& {Ott}(2013)}]{couch_2013_aa}
{Couch}, S.~M., \& {Ott}, C.~D. 2013, \apjl, 778, L7,
  \dodoi{10.1088/2041-8205/778/1/L7}

\bibitem[{{Couch} \& {Ott}(2015)}]{couch_2015_ab}
---. 2015, \apj, 799, 5, \dodoi{10.1088/0004-637X/799/1/5}

\bibitem[{{Curtis} {et~al.}(2020){Curtis}, {Wolfe}, {Fr{\"o}hlich}, {Miller},
  {Wollaeger}, \& {Ebinger}}]{curtis_2020_aa}
{Curtis}, S., {Wolfe}, N., {Fr{\"o}hlich}, C., {et~al.} 2020, arXiv e-prints,
  arXiv:2008.05498.
\newblock \doarXiv{2008.05498}

\bibitem[{Dubey {et~al.}(2009)Dubey, Antypas, Ganapathy, Reid, Riley, Sheeler,
  Siegel, \& Weide}]{dubey_2009_aa}
Dubey, A., Antypas, K., Ganapathy, M.~K., {et~al.} 2009, Parallel Computing,
  35, 512 , \dodoi{https://doi.org/10.1016/j.parco.2009.08.001}

\bibitem[{{Farmer} {et~al.}(2016){Farmer}, {Fields}, {Petermann}, {Dessart},
  {Cantiello}, {Paxton}, \& {Timmes}}]{farmer_2016_aa}
{Farmer}, R., {Fields}, C.~E., {Petermann}, I., {et~al.} 2016, \apjs, 227, 22,
  \dodoi{10.3847/1538-4365/227/2/22}

\bibitem[{{Fields} \& {Couch}(2020)}]{fields_2020_aa}
{Fields}, C.~E., \& {Couch}, S.~M. 2020, \apj, 901, 33,
  \dodoi{10.3847/1538-4357/abada7}

\bibitem[{{Fields} {et~al.}(2018){Fields}, {Timmes}, {Farmer}, {Petermann},
  {Wolf}, \& {Couch}}]{fields_2018_aa}
{Fields}, C.~E., {Timmes}, F.~X., {Farmer}, R., {et~al.} 2018, \apjs, 234, 19,
  \dodoi{10.3847/1538-4365/aaa29b}

\bibitem[{{Foglizzo} {et~al.}(2006){Foglizzo}, {Scheck}, \&
  {Janka}}]{foglizzo_2006_aa}
{Foglizzo}, T., {Scheck}, L., \& {Janka}, H.~T. 2006, \apj, 652, 1436,
  \dodoi{10.1086/508443}

\bibitem[{{Fryer} {et~al.}(2019){Fryer}, {Lloyd-Ronning}, {Wollaeger},
  {Wiggins}, {Miller}, {Dolence}, {Ryan}, \& {Fields}}]{fryer_2019_aa}
{Fryer}, C.~L., {Lloyd-Ronning}, N., {Wollaeger}, R., {et~al.} 2019, European
  Physical Journal A, 55, 132, \dodoi{10.1140/epja/i2019-12818-y}

\bibitem[{{Fryxell} {et~al.}(2000){Fryxell}, {Olson}, {Ricker}, {Timmes},
  {Zingale}, {Lamb}, {MacNeice}, {Rosner}, {Truran}, \&
  {Tufo}}]{fryxell_2000_aa}
{Fryxell}, B., {Olson}, K., {Ricker}, P., {et~al.} 2000, \apjs, 131, 273,
  \dodoi{10.1086/317361}

\bibitem[{{Herwig} {et~al.}(2006){Herwig}, {Freytag}, {Hueckstaedt}, \&
  {Timmes}}]{herwig_2006_aa}
{Herwig}, F., {Freytag}, B., {Hueckstaedt}, R.~M., \& {Timmes}, F.~X. 2006,
  \apj, 642, 1057, \dodoi{10.1086/501119}

\bibitem[{Hunter(2007)}]{hunter_2007_aa}
Hunter, J.~D. 2007, Computing In Science \&amp; Engineering, 9, 90

\bibitem[{{Jones} {et~al.}(2017){Jones}, {Andrassy}, {Sandalski}, {Davis},
  {Woodward}, \& {Herwig}}]{jones_2017_aa}
{Jones}, S., {Andrassy}, R., {Sandalski}, S., {et~al.} 2017, \mnras, 465, 2991,
  \dodoi{10.1093/mnras/stw2783}

\bibitem[{{Langanke} \& {Mart{\'{\i}}nez-Pinedo}(2000)}]{langanke_2000_aa}
{Langanke}, K., \& {Mart{\'{\i}}nez-Pinedo}, G. 2000, Nuclear Physics A, 673,
  481, \dodoi{10.1016/S0375-9474(00)00131-7}

\bibitem[{{Lattimer} \& {Swesty}(1991)}]{lattimer_1991_aa}
{Lattimer}, J.~M., \& {Swesty}, D.~F. 1991, \nphysa, 535, 331,
  \dodoi{10.1016/0375-9474(91)90452-C}

\bibitem[{Lee \& Deane(2009)}]{lee_2008_aa}
Lee, D., \& Deane, A.~E. 2009, Journal of Computational Physics, 228, 952 ,
  \dodoi{https://doi.org/10.1016/j.jcp.2008.08.026}

\bibitem[{{Meakin} \& {Arnett}(2007)}]{meakin_2007_ab}
{Meakin}, C.~A., \& {Arnett}, D. 2007, \apj, 667, 448

\bibitem[{{M{\"o}sta} {et~al.}(2015){M{\"o}sta}, {Ott}, {Radice}, {Roberts},
  {Schnetter}, \& {Haas}}]{mosta_2015_aa}
{M{\"o}sta}, P., {Ott}, C.~D., {Radice}, D., {et~al.} 2015, \nat, 528, 376,
  \dodoi{10.1038/nature15755}

\bibitem[{{M{\"u}ller} \& {Janka}(2015)}]{muller_2015_ab}
{M{\"u}ller}, B., \& {Janka}, H.~T. 2015, \mnras, 448, 2141,
  \dodoi{10.1093/mnras/stv101}

\bibitem[{{M{\"u}ller} {et~al.}(2017){M{\"u}ller}, {Melson}, {Heger}, \&
  {Janka}}]{muller:2017}
{M{\"u}ller}, B., {Melson}, T., {Heger}, A., \& {Janka}, H.-T. 2017, \mnras,
  472, 491

\bibitem[{{M{\"u}ller} \& {Varma}(2020)}]{muller_2020_aa}
{M{\"u}ller}, B., \& {Varma}, V. 2020, \mnras, 498, L109,
  \dodoi{10.1093/mnrasl/slaa137}

\bibitem[{{M{\"u}ller} {et~al.}(2016){M{\"u}ller}, {Viallet}, {Heger}, \&
  {Janka}}]{muller_2016_aa}
{M{\"u}ller}, B., {Viallet}, M., {Heger}, A., \& {Janka}, H.-T. 2016, ArXiv
  e-prints.
\newblock \doarXiv{1605.01393}

\bibitem[{{O'Connor} \& {Ott}(2011)}]{oconnor_2011_aa}
{O'Connor}, E., \& {Ott}, C.~D. 2011, \apj, 730, 70,
  \dodoi{10.1088/0004-637X/730/2/70}

\bibitem[{{O'Connor} \& {Couch}(2018{\natexlab{a}})}]{oconnor_2018_ab}
{O'Connor}, E.~P., \& {Couch}, S.~M. 2018{\natexlab{a}}, \apj, 865, 81,
  \dodoi{10.3847/1538-4357/aadcf7}

\bibitem[{{O'Connor} \& {Couch}(2018{\natexlab{b}})}]{oconnor_2018_aa}
---. 2018{\natexlab{b}}, \apj, 854, 63, \dodoi{10.3847/1538-4357/aaa893}

\bibitem[{{Pajkos} {et~al.}(2021){Pajkos}, {Warren}, {Couch}, {O'Connor}, \&
  {Pan}}]{pajkos_2021_aa}
{Pajkos}, M.~A., {Warren}, M.~L., {Couch}, S.~M., {O'Connor}, E.~P., \& {Pan},
  K.-C. 2021, \apj, 914, 80, \dodoi{10.3847/1538-4357/abfb65}

\bibitem[{{Paxton} {et~al.}(2011){Paxton}, {Bildsten}, {Dotter}, {Herwig},
  {Lesaffre}, \& {Timmes}}]{paxton_2011_aa}
{Paxton}, B., {Bildsten}, L., {Dotter}, A., {et~al.} 2011, \apjs, 192, 3,
  \dodoi{10.1088/0067-0049/192/1/3}

\bibitem[{{Paxton} {et~al.}(2013){Paxton}, {Cantiello}, {Arras}, {Bildsten},
  {Brown}, {Dotter}, {Mankovich}, {Montgomery}, {Stello}, {Timmes}, \&
  {Townsend}}]{paxton_2013_aa}
{Paxton}, B., {Cantiello}, M., {Arras}, P., {et~al.} 2013, \apjs, 208, 4,
  \dodoi{10.1088/0067-0049/208/1/4}

\bibitem[{{Paxton} {et~al.}(2015){Paxton}, {Marchant}, {Schwab}, {Bauer},
  {Bildsten}, {Cantiello}, {Dessart}, {Farmer}, {Hu}, {Langer}, {Townsend},
  {Townsley}, \& {Timmes}}]{paxton_2015_aa}
{Paxton}, B., {Marchant}, P., {Schwab}, J., {et~al.} 2015, \apjs, 220, 15,
  \dodoi{10.1088/0067-0049/220/1/15}

\bibitem[{{Paxton} {et~al.}(2018){Paxton}, {Schwab}, {Bauer}, {Bildsten},
  {Blinnikov}, {Duffell}, {Farmer}, {Goldberg}, {Marchant}, {Sorokina},
  {Thoul}, {Townsend}, \& {Timmes}}]{paxton_2018_aa}
{Paxton}, B., {Schwab}, J., {Bauer}, E.~B., {et~al.} 2018, \apjs, 234, 34,
  \dodoi{10.3847/1538-4365/aaa5a8}

\bibitem[{{Paxton} {et~al.}(2019){Paxton}, {Smolec}, {Gautschy}, {Bildsten},
  {Cantiello}, {Dotter}, {Farmer}, {Goldberg}, {Jermyn}, {Kanbur}, {Marchant},
  {Schwab}, {Thoul}, {Townsend}, {Wolf}, {Zhang}, \& {Timmes}}]{paxton_2019_aa}
{Paxton}, B., {Smolec}, R., {Gautschy}, A., {et~al.} 2019, arXiv e-prints.
\newblock \doarXiv{1903.01426}

\bibitem[{{Pejcha} \& {Thompson}(2015)}]{pejcha_2015_ab}
{Pejcha}, O., \& {Thompson}, T.~A. 2015, \apj, 801, 90,
  \dodoi{10.1088/0004-637X/801/2/90}

\bibitem[{{Ritter} {et~al.}(2018){Ritter}, {Andrassy}, {C{\^o}t{\'e}},
  {Herwig}, {Woodward}, {Pignatari}, \& {Jones}}]{ritter_2018_aa}
{Ritter}, C., {Andrassy}, R., {C{\^o}t{\'e}}, B., {et~al.} 2018, \mnras, 474,
  L1, \dodoi{10.1093/mnrasl/slx126}

\bibitem[{Schaeffer(2013)}]{sht_2013_aa}
Schaeffer, N. 2013, Geochemistry, Geophysics, Geosystems, 14, 751,
  \dodoi{10.1002/ggge.20071}

\bibitem[{{Sukhbold} {et~al.}(2016){Sukhbold}, {Ertl}, {Woosley}, {Brown}, \&
  {Janka}}]{sukhbold_2016_aa}
{Sukhbold}, T., {Ertl}, T., {Woosley}, S.~E., {Brown}, J.~M., \& {Janka}, H.-T.
  2016, \apj, 821, 38, \dodoi{10.3847/0004-637X/821/1/38}

\bibitem[{{Timmes} {et~al.}(2000){Timmes}, {Hoffman}, \&
  {Woosley}}]{timmes_2000_ab}
{Timmes}, F.~X., {Hoffman}, R.~D., \& {Woosley}, S.~E. 2000, \apjs, 129, 377

\bibitem[{{Timmes} \& {Swesty}(2000)}]{timmes_2000_aa}
{Timmes}, F.~X., \& {Swesty}, F.~D. 2000, \apjs, 126, 501

\bibitem[{Toro(1999)}]{toro_1999_aa}
Toro, E.~F. 1999, Riemann Solvers and Numerical Methods for Fluid Dynamics
  (Springer, Berlin, Heidelberg)

\bibitem[{{Trampedach} {et~al.}(2014){Trampedach}, {Stein},
  {Christensen-Dalsgaard}, {Nordlund}, \& {Asplund}}]{trampedach_2014_aa}
{Trampedach}, R., {Stein}, R.~F., {Christensen-Dalsgaard}, J., {Nordlund},
  {\AA}., \& {Asplund}, M. 2014, \mnras, 445, 4366,
  \dodoi{10.1093/mnras/stu2084}

\bibitem[{{Turk} {et~al.}(2011){Turk}, {Smith}, {Oishi}, {Skory}, {Skillman},
  {Abel}, \& {Norman}}]{turk_2011_aa}
{Turk}, M.~J., {Smith}, B.~D., {Oishi}, J.~S., {et~al.} 2011, \apjs, 192, 9,
  \dodoi{10.1088/0067-0049/192/1/9}

\bibitem[{{Varma} \& {M{\"u}ller}(2021)}]{varma_2021_aa}
{Varma}, V., \& {M{\"u}ller}, B. 2021, arXiv e-prints, arXiv:2101.00213.
\newblock \doarXiv{2101.00213}

\bibitem[{Warren {et~al.}(2020)Warren, Couch, O'Connor, \&
  Morozova}]{warren_2020_aa}
Warren, M.~L., Couch, S.~M., O'Connor, E.~P., \& Morozova, V. 2020, The
  Astrophysical Journal, 898, 139, \dodoi{10.3847/1538-4357/ab97b7}

\bibitem[{{Yadav} {et~al.}(2019){Yadav}, {M{\"u}ller}, {Janka}, {Melson}, \&
  {Heger}}]{yadav_2019_aa}
{Yadav}, N., {M{\"u}ller}, B., {Janka}, H.~T., {Melson}, T., \& {Heger}, A.
  2019, arXiv e-prints, arXiv:1905.04378.
\newblock \doarXiv{1905.04378}

\bibitem[{Yoshida {et~al.}(2019)Yoshida, Takiwaki, Kotake, Takahashi, Nakamura,
  \& Umeda}]{Yoshida_2019}
Yoshida, T., Takiwaki, T., Kotake, K., {et~al.} 2019, The Astrophysical
  Journal, 881, 16, \dodoi{10.3847/1538-4357/ab2b9d}

\bibitem[{{Yoshida} {et~al.}(2021){Yoshida}, {Takiwaki}, {Kotake}, {Takahashi},
  {Nakamura}, \& {Umeda}}]{yoshida_2021_aa}
{Yoshida}, T., {Takiwaki}, T., {Kotake}, K., {et~al.} 2021, \apj, 908, 44,
  \dodoi{10.3847/1538-4357/abd3a3}

\bibitem[{{Zingale} {et~al.}(2002){Zingale}, {Dursi}, {ZuHone}, {Calder},
  {Fryxell}, {Plewa}, {Truran}, {Caceres}, {Olson}, {Ricker}, {Riley},
  {Rosner}, {Siegel}, {Timmes}, \& {Vladimirova}}]{zingale_2002_aa}
{Zingale}, M., {Dursi}, L.~J., {ZuHone}, J., {et~al.} 2002, \apjs, 143, 539,
  \dodoi{10.1086/342754}

\end{thebibliography}

\end{document}